%% file: ms.tex
\documentclass[fapj]{emulateapj} 
\usepackage{amsmath}
\usepackage{natbib}
\usepackage{graphicx}
\usepackage{epsf}
\usepackage{color}
\usepackage{epsfig}
\usepackage{multirow}
\DeclareGraphicsExtensions{.jpg,.pdf,.png,.eps,.ps}

\input{definitions.tex}

\begin{document}

\submitted{\today.} 

\title{Cosmic microwave background constraints on the duration and timing of reionization from the South Pole Telescope}
\author{
 O.~Zahn,\altaffilmark{1} 
 C.~L.~Reichardt,\altaffilmark{2}
  L.~Shaw,\altaffilmark{3} 
  A.~Lidz,\altaffilmark{4}
K.~A.~Aird,\altaffilmark{5}
 B.~A.~Benson,\altaffilmark{6,7}
 L.~E.~Bleem,\altaffilmark{6,8}
 J.~E.~Carlstrom,\altaffilmark{6,7,8,9,10}
 C.~L.~Chang,\altaffilmark{6,7,10}
 H.~M. Cho, \altaffilmark{11}
 T.~M.~Crawford,\altaffilmark{6,9}
 A.~T.~Crites,\altaffilmark{6,9}
 T.~de~Haan,\altaffilmark{12}
 M.~A.~Dobbs,\altaffilmark{12}
 O.~Dor\'e,\altaffilmark{17}
  J.~Dudley,\altaffilmark{12}
 E.~M.~George,\altaffilmark{3}
 N.~W.~Halverson,\altaffilmark{13}
 G.~P.~Holder,\altaffilmark{12}
 W.~L.~Holzapfel,\altaffilmark{3}
 S.~Hoover,\altaffilmark{6,8}
 Z.~Hou,\altaffilmark{14}
 J.~D.~Hrubes,\altaffilmark{5}
 M.~Joy,\altaffilmark{15}
  R.~Keisler,\altaffilmark{6,8}
 L.~Knox,\altaffilmark{14}
 A.~T.~Lee,\altaffilmark{3,16}
 E.~M.~Leitch,\altaffilmark{6,9}
 M.~Lueker,\altaffilmark{17}
 D.~Luong-Van,\altaffilmark{5}
 J.~J.~McMahon,\altaffilmark{18}
 J.~Mehl,\altaffilmark{6}
 S.~S.~Meyer,\altaffilmark{6,7,8,9}
 M.~Millea,\altaffilmark{14}
 J.~J.~Mohr,\altaffilmark{19,20,21}
 T.~E.~Montroy,\altaffilmark{22}
 T.~Natoli,\altaffilmark{6,8}
 S.~Padin,\altaffilmark{6,9,17}
 T.~Plagge,\altaffilmark{6,9}
 C.~Pryke,\altaffilmark{6,7,9,23}
 J.~E.~Ruhl,\altaffilmark{22}
 K.~K.~Schaffer,\altaffilmark{6,7,24}
 E.~Shirokoff,\altaffilmark{3} 
 H.~G.~Spieler,\altaffilmark{16}
 Z.~Staniszewski,\altaffilmark{23}
 A.~A.~Stark,\altaffilmark{25}
 K.~Story,\altaffilmark{6,8}
 A.~van~Engelen,\altaffilmark{12}
 K.~Vanderlinde,\altaffilmark{12}
 J.~D.~Vieira,\altaffilmark{17} and 
 R.~Williamson\altaffilmark{6,9} }

\altaffiltext{1}{Berkeley Center for Cosmological Physics,
Department of Physics, University of California, and Lawrence Berkeley
National Labs, Berkeley, CA, USA 94720}
\altaffiltext{2}{Department of Physics,
University of California, Berkeley, CA, USA 94720}
\altaffiltext{3}{Department of Physics, Yale University, P.O. Box 208210, New Haven,
CT, USA 06520-8120}
\altaffiltext{4}{Department of Physics \& Astronomy, University of Pennsylvania, 209 South 33rd Street, Philadelphia, PA 19104}

\altaffiltext{5}{University of Chicago,
5640 South Ellis Avenue, Chicago, IL, USA 60637}
\altaffiltext{6}{Kavli Institute for Cosmological Physics,
University of Chicago, 5640 South Ellis Avenue, Chicago, IL, USA 60637}
\altaffiltext{7}{Enrico Fermi Institute,
University of Chicago,
5640 South Ellis Avenue, Chicago, IL, USA 60637}
\altaffiltext{8}{Department of Physics,
University of Chicago,
5640 South Ellis Avenue, Chicago, IL, USA 60637}
\altaffiltext{9}{Department of Astronomy and Astrophysics,
University of Chicago,
5640 South Ellis Avenue, Chicago, IL, USA 60637}
\altaffiltext{10}{Argonne National Laboratory, 9700 S. Cass Avenue, Argonne, IL, USA 60439}
\altaffiltext{11}{NIST Quantum Devices Group, 325 Broadway Mailcode 817.03, Boulder, CO, USA 80305}
\altaffiltext{12}{Department of Physics,
McGill University, 3600 Rue University, 
Montreal, Quebec H3A 2T8, Canada}
\altaffiltext{13}{Department of Astrophysical and Planetary Sciences and Department of Physics,
University of Colorado,
Boulder, CO, USA 80309}
\altaffiltext{14}{Department of Physics, 
University of California, One Shields Avenue, Davis, CA, USA 95616}
\altaffiltext{15}{Department of Space Science, VP62,
NASA Marshall Space Flight Center,
Huntsville, AL, USA 35812}
\altaffiltext{16}{Physics Division,
Lawrence Berkeley National Laboratory,
Berkeley, CA, USA 94720}
\altaffiltext{17}{California Institute of Technology, MS 249-17, 1216 E. California Blvd., Pasadena, CA, USA 91125}
\altaffiltext{18}{Department of Physics, University of Michigan, 450 Church Street, Ann  
Arbor, MI, USA 48109}
\altaffiltext{19}{Department of Physics,
Ludwig-Maximilians-Universit\"{a}t,
Scheinerstr.\ 1, 81679 M\"{u}nchen, Germany}
\altaffiltext{20}{Excellence Cluster Universe,
Boltzmannstr.\ 2, 85748 Garching, Germany}
\altaffiltext{21}{Max-Planck-Institut f\"{u}r extraterrestrische Physik,
Giessenbachstr.\ 85748 Garching, Germany}
\altaffiltext{22}{Physics Department, Center for Education and Research in Cosmology 
and Astrophysics, 
Case Western Reserve University,
Cleveland, OH, USA 44106}
\altaffiltext{23}{Department of Physics, University of Minnesota, 116 Church Street S.E. Minneapolis, MN, USA 55455}
\altaffiltext{24}{Liberal Arts Department, 
School of the Art Institute of Chicago, 
112 S Michigan Ave, Chicago, IL, USA 60603}
\altaffiltext{25}{Harvard-Smithsonian Center for Astrophysics,
60 Garden Street, Cambridge, MA, USA 02138}


\email{zahn@berkeley.edu}

\begin{abstract}

The epoch of reionization is a milestone of cosmological structure
formation, marking the birth of the first objects massive enough to
yield large numbers of ionizing photons. However, the mechanism and
timescale of reionization remain largely unknown.  Measurements of the
CMB Doppler effect from ionizing bubbles embedded in large-scale
velocity streams -- known as the patchy kinetic Sunyaev-Zel'dovich
(SZ) effect -- can be used to constrain the duration of
reionization. When combined with large-scale CMB polarization
measurements, the evolution of the ionized fraction, $\xhii$, can be
inferred. 
Using new multi-frequency data from the
South Pole Telescope (SPT), we show that the ionized fraction evolved
relatively rapidly.  For our basic foreground model, we find the
kinetic SZ power sourced by reionization at $\ell=3000$ to be
$D_{3000}^{\rm patchy}\leq 2.1\, \mu K^2$ at 95\% confidence.  Using
reionization simulations, we translate this to a limit on the duration
of reionization of $\Delta z \equiv z_{\xhii=0.20}-z_{\xhii=0.99} \le
4.4$ (95\% confidence).  We find that this constraint depends on
assumptions about the angular correlation between the thermal SZ power
and the cosmic infrared background (CIB).  Introducing the degree of
correlation as a free parameter, we find that the limit on kSZ power
weakens to $D_{3000}^{\rm patchy}\leq 4.9\, \mu K^2$, implying $\Delta
z \le 7.9$ (95\% confidence).
We combine the SPT constraint on the duration of reionization with the WMAP measurement of the integrated optical depth to probe the cosmic ionization history. 
We find that reionization ended with 95\% confidence at $z > 7.2$ under the 
assumption of no tSZ-CIB correlation,
and $z>5.8$ when correlations are allowed.
Improved constraints from the full SPT data set in conjunction with upcoming \herschel\ and \planck\ data should detect extended reionization at $>$\,95\% confidence provided $\Delta z\geq 2$. 
These CMB observations complement other observational probes of the epoch of reionization such as the redshifted 21 cm line and and narrow-band surveys for Lyman-$\alpha$ emitting galaxies. 

\end{abstract}


\section{Introduction}\label{sec:intro}
\setcounter{footnote}{0}

Galaxy-sized dark matter halos first collapse at $z\geq 25$.  
The stars and black holes that form in these halos ultimately ionize and heat the intergalactic medium (IGM). However, exactly when and how this process occurred is unknown. 

To date, there have been two primary observational constraints on the reionization era.  First, Lyman-$\alpha$ (Ly$\alpha$) forest absorption spectra towards high redshift quasars show that the opacity of the IGM to Ly$\alpha$ photons is rapidly increasing at $z\gsim6$ \citep[e.g.,][]{Fan2006}.
This increase in opacity has been interpreted as evidence for an increasing neutral hydrogen fraction. 
However, the interpretation of quasar absorption spectra is hampered by the large cross section for Ly$\alpha$ absorption, which can lead to complete absorption even if the hydrogen ionization fraction is high. 
At the very least, the amount of transmission in the Ly$\alpha$ forest at $z\lesssim6$ indicates that the {\it bulk} of reionization occurred at higher redshifts \citep{McGreer2011}. 

Second, WMAP measurements of the optical depth through large-angle cosmic microwave background (CMB) polarization anisotropy suggest that the redshift of reionization assuming an instantaneous process is $z\simeq 10.6 \pm 1.4$ \citep{komatsu11}. 
CMB polarization measurements offer an integral constraint on the reionization history \citep{Kogut2003, Page2007}. 
Therefore, the polarization data are fully consistent with either instantaneous or extended reionization  scenarios. 

Together these observations have been interpreted by some as favoring a prolonged reionization epoch ending at $z\simeq 6$ (\citealt{MiraldaEscude:2002yd,Cen2003, Fan2006, BoltonHaehnelt2007b, WyitheCen2007}). However, the claim that $z \sim 6$ Ly-$\alpha$ forest spectra
probe the tail end of reionization is somewhat controversial. These data may be consistent with reionization completing at a redshift higher than $z=6$ (e.g., \citealt{Oh:2004rm,LidzOF2006,BeckerRS2007,McGreer2011}).

Additional model-dependent constraints on reionization, most of them {\it upper} limits on the neutral fraction, have been derived by other means: (1) the size of the proximity zone around quasars (\citealt{WyitheLC2005, Fan2006}, but see \citealt{MesingerHC2004, BoltonHaehnelt2007a, Lidz2007, Maselli2007}); 
(2) the claimed detections of damping wing absorption from neutral IGM in quasar spectra \citep{MesingerHaiman2004, MesingerHaiman2007} 
which were used to place {\it lower} limits of 20\% and 3\% on the neutral fraction at z=6.3 (but see \citealt{MesingerFurlanetto2008a}), and more recently a lower limit of $10\%$ at $z=7.1$ \citep{Mortlock:2011va}; (3) the {\it non}-detection of intergalactic damping wing absorption in the spectrum of a gamma ray burst at z=6.3 \citep{Totani2006,  McQuinn2008}; 
and (4) the number density and clustering of Ly$\alpha$ emitters \citep{MalhotraRhoads2004, HaimanCen2005, FurlanettoZH2006, Kashikawa2006, McQuinn2007a, MesingerFurlanetto2008b,Ouchi2010}.  These constraints will be discussed in more detail in \S\ref{sec:discussion} and Figure \ref{fig:summary}.

In this paper, we produce the first constraint on the evolution of the ionized fraction using CMB data. 
We use new data from the South Pole Telescope (SPT) to constrain the amplitude of the ``patchy'' kinetic Sunyaev-Zel'dovich (kSZ) signal resulting from inhomogeneous reionization \citep{knox98,Gruzinov:1998u,Santos:2003jb,zahn05,McQuinn2005,iliev06,fan06}. We interpret this signal in the context of reionization models to place the first constraints on the duration of the epoch of reionization using CMB data.
A companion paper \citep[hereafter R11]{reichardt11} describes  the SPT data underlying this measurement in detail. 
We combine the SPT data with WMAP7 measurements of the integrated opacity due to reionization. 
The combination makes it possible to constrain the evolution of the ionized fraction, significantly limiting the allowed range of reionization scenarios.

The structure of the paper is as follows. 
In \S\ref{sec:pkszmodeling} we outline the physics of the kSZ effect from inhomogeneous reionization (``patchy kSZ'') and introduce our simulation scheme. 
 We describe the data in \S\ref{sec:data}. 
In \S\ref{sec:fitting}, we set up the model for the CMB and astronomical foregrounds which we fit to the data. 
In \S\ref{sec:tauksz}, we review the constraint on the integrated optical depth by WMAP7 and present the SPT constraints on the amplitude of the patchy kSZ component. 
In \S\ref{sec:results}, we present our constraints on the evolution of the ionized fraction, and forecast future constraints with the SPT, \planck, and \herschel\ experiments. 
 We summarize our results and place them in the context of other observations in \S\ref{sec:discussion}.

\section{Epoch of reionization and  kinetic Sunyaev-Zel'dovich effect }
\label{sec:pkszmodeling}

In this section, we review the theory of the kSZ signal, and then introduce the reionization simulations and discuss the kinetic SZ power used to fit to the data.

\subsection{CMB reionization observables}

Scattering of CMB photons into the line-of-sight introduces
temperature anisotropy through the Doppler effect if there are
perturbations in the baryon density, $\rho_{\rm b}$, or ionization
fraction, $x_e$. These scatterings slightly change the temperature of
the CMB blackbody.  The CMB Doppler effect is often called the kinetic
Sunyaev-Zel'dovich (kSZ) effect \citep{sunyaev70b,sunyaev80} when
referring to reionization or the nonlinear regime of structure
formation, and the Ostriker-Vishniac effect \citep{ostriker86} when
referring to the linear regime, but we will call it kSZ effect throughout. The total contribution to the temperature
anisotropy from a redshift interval $[z_1,z_2]$ is given by:
\beq
\frac{\Delta T_{\rm kSZ}}{T_{\rm CMB}} (\n) = \frac{\sigma_T}{c}  \int_{z_1}^{z_2}\frac{dx}{dz}\frac{dz}{(1+z)} \bar{n}_e(z) e^{-\tau(z)}\n \cdot \mathbf{q} ,
\eeq
where $\sigma_T$ is the Thomson scattering cross-section, $\frac{dx}{dz}$ is the comoving line-element and $\n$ is the line of sight unit vector. $\bar{n}_e(z)$ is the mean free electron density,  
\beq
{\bar n}_e(z) = \frac{\bar{x}_e(z)\bar{\rho}_b(z)}{\mu_e m_p} \, ,
\eeq
where $\bar{x}_e(z)$ and $\bar{\rho}_b(z)$ are the mean free electron fraction and mean baryon density of the universe as functions of redshift and $\mu_e m_p$ is the mean mass per electron. We set $\mu_e = 1.22$, appropriate for singly ionized helium. Due to their similar ionization potential, we assume that helium is singly ionized alongside hydrogen, that is, when all hydrogen is ionized and all helium is singly ionized $\xhii=1$. Note that $\xhii = 1.07$ when helium is fully ionized; our model for the post-reionization kinetic SZ signal assumes that this occurs at $z = 3$. 

The optical depth, $\tau(z)$, from the observer to redshift $z$ is given by, 
 \beq
\tau(z)=\sigma_T \int_0^z \frac{{\bar n}_e(z^\prime)}{1+z^\prime}\frac{dx}{dz^\prime}dz^\prime .
 \label{eq:tau}
  \eeq
The \wmap\ large-scale polarization signal constrains the optical depth to the end of recombination, $\tau(z \sim 1000) = 0.088 \pm 0.014$.

Finally, inhomogeneities in the ionization fraction and baryon density enter through
\beq
\mathbf{q}=\left(1+\delta_x\right) \left(1+\delta_{\rm b}\right) \mathbf{v}
\, .
\label{eq:ksz_source}
\eeq 
where $\delta_x=x_e/{\xhii}-1$, $\delta_b=\rho_b/{\bar \rho_b}-1$, and $\mathbf{v}$ is the bulk motion of free electrons  with respect to the CMB. 

We discuss the post-reionization kSZ
signal in \S\ref{sec:homog_ksz}. Reionization produces kSZ power
primarily through local changes in the ionization fraction. To predict
the amplitude of this signal we require a prescription for calculating
the evolution of the ionization morphology during the epoch of reionization. 

 \subsection{Efficient Monte Carlo reionization simulations}
 \label{sec:reisim}
 
 Our ability to infer  the properties of the first ionizing sources from observations
hinges on the accuracy with which the ionization morphology during reionization can be modeled.
 A number of groups have developed 3-D radiative transfer codes (e.g., \citealt{Gnedin2000, Sokasian2001, Razoumov2002, Ciardi2003, Mellema2006, McQuinn2007b, SemelinCB2007, TracCen2007,  AltayCP2008, AubertTeyssier2008, FinlatorOD2009, PetkovaSpringel2009}). 
However, the mass resolution and volume requirements for simulations of reionization are daunting.  
Simulations must resolve the low-mass galaxies (down to the atomic cooling threshold corresponding to a host-halo mass $\sim 10^8\;M_\odot$ at $z\sim$7--10) that are expected to dominate the ionizing photon budget. 
They must also be large enough to statistically sample the distribution of ionized regions. 
These regions can span tens of comoving Mpc in size towards the end of reionization
\citep{FurlanettoOh2005, Zahn2005, Zahn2007, MesingerFurlanetto2007, ShinTC2008}. 
Recently, some groups have  come close to achieving this dynamic range in a single simulation  (\citealt{Iliev2006b, McQuinn2007b, TracCen2007, ShinTC2008, TracCL2008,Zahn2010}; see  \citealt{TracGnedin2009} for a recent review).\footnote{Note however that Lyman limit systems (LLSs), which can dominate the absorption of ionizing photons (see for example the appendix of \citealt{FurlanettoOh2005}), are still too small to be resolved by state-of-the-art reionization simulations, and are typically included via analytic prescriptions (e.g., \citealt{FurlanettoOh2005,McQuinn2007b,ChoudhuryHR2009,Crociani:2010qe}).}  

Given the large uncertainties in the production and escape rates of ionizing photons in high-redshift galaxies (and also in how the photons are absorbed by dense systems), a large parameter space must be explored to interpret observations.  
These concerns have prompted several groups to develop much more CPU efficient, approximate algorithms \citep{Furlanetto2004,Zahn2005,Zahn2007,MesingerFurlanetto2007,GeilWyithe2008, Alvarez2009, ChoudhuryHR2009, Thomas2009,Zahn2010}. 
The basic idea is that semi-analytic models 
can be applied in a Monte Carlo fashion to large-scale realizations of the density and velocity fields. 
The resulting complex reionization morphology can be compared side-by-side with simulations based on the same initial conditions combined with the same source and sink prescriptions. 
 We use a modification of the \citet{Zahn2010} model to set up our reionization parameter space.\footnote{The reason we do not use purely analytic models is that they have difficulties describing the  partial overlap stage of merging ionized regions (e.g., \citealt{McQuinn:2005ce}).} 

The model associates ionized regions with the ionizing sources they contain. 
The size distribution of ionized regions is related to the halo mass function through the ansatz, $M_{\rm ion}=\zeta M_{\rm gal}$. 
 $M_{\rm gal}$ is the mass in collapsed objects. 
$\zeta$ is the efficiency factor for ionization, and can be decomposed as $\zeta=f_{\rm esc}f_* N_{\gamma/b} n_{\rm rec}^{-1}$.
 Here, $f_{\rm esc}$ is the escape fraction of ionizing photons from the object, $f_*$ is the star formation efficiency, $N_{\gamma/b}$ the number of ionizing photons produced per baryon converted into stars, and $n_{\rm rec}$ is the typical number of times a hydrogen atom has recombined. 
Although the efficiency factor is a rough combination of uncertain source properties, it can encapsulate a wide variety of reionization scenarios. 

We calculate $M_{\rm gal}$ according to the extended Press-Schechter model \citep{Bond:1990iw}. 
In this model, the collapsed fraction (or the fraction of baryons that lie in galaxies) in a region of size $r$ depends on the mean overdensity of that region, $\overline{\delta}(r)$, as: 
\begin{equation}
f_{\rm coll}(r,M_{\rm min})=\erfc\left[\frac{\delta_c(z)-\overline{\delta}(r)}{\sqrt{2[\sigma^2(r_{\rm min})-\sigma^2(r)]}}\right]
\, .
\label{collfract} 
\end{equation}
Here, $\delta_c(z)$ is a numerical factor from linear theory equal to $1.686$ today, and $\sigma^2(r)$ is the linear theory rms fluctuation smoothed on scale $r$. 
$r_{\rm min}$ is defined as the radius that encloses the mass $M_{\rm min}$ (at average density $\overline{\rho}$) corresponding to a virial temperature of $10^4\,$K, above which atomic hydrogen line cooling becomes efficient. 
The redshift dependence of the minimum mass (see \citealt{Barkana:2000fd}) is given in terms of the virial temperature: 
\begin{equation}
M_{\rm min} \simeq 10^8 \frac{M_\odot}{h}
\left[\frac{T_{\rm vir}}{2 \times 10^{4} K} \frac{10}{1+z} \right]^{3/2}
\left[\frac{\Omega_0}{\Omega_m(z)} \frac{\Delta_c(z)}{18 \pi^2}\right]^{-1/2}
\end{equation}
with the fitting function $\Delta_c(z)=18 \pi^2 + 82 d - 39 d^2$ \citep{Bryan:1997dn}, 
where $d\equiv \Omega_m (z)-1$ is evaluated at the collapse redshift. 
The luminous mass in galaxies required to fully ionize all hydrogen atoms is inversely proportional to the ionizing efficiency, so ionization requires  
\begin{equation}
f_{\rm coll} \geq \zeta^{-1} \, .
\end{equation}
Hence, we can define a barrier $\delta_x(r,z)$ which fluctuations have to cross for their baryonic content to become ionized:
\begin{equation}
\delta(r) \geq \delta_x(r,z) \equiv \delta_c(z)-\sqrt{2}
\erfc^{-1}(\zeta^{-1}) [\sigma^2(r_{\rm min})-\sigma^2(r)]^{1/2} \, .
\label{eq:barrier} 
\end{equation}

The algorithm then proceeds as follows. For every position, the
linear matter over-density is calculated within a range given by the smoothing kernel (initially set to a large value comparable to the simulation size) . \cite{Zahn2010}  show that good agreement with radiative transfer simulations is achieved when the smoothing kernel is a top hat in harmonic space.  
The halo collapse fraction is estimated from this over-density, and
translated with the efficiency factor $\zeta$ into an expected number of ionizing photons based on the
number of collapsed halos above the atomic cooling threshold mass.
The number of ionizing photons is then compared to the number of hydrogen atoms
at each point.
If there are sufficient ionizing photons, a point is labeled ionized.  If not, a smaller radius is set
and the algorithm repeated until the resolution of the simulation box is reached.

We optimize the algorithm for mock patchy kSZ spectra at the SPT angular  resolution. 
First, we simulate the density and velocity fields in a rectangular box with a high dynamic range, $1^2 \times 3$ Gpc/h with $512^2 \times 1536$ volume elements. 
The simulation box subtends 8.5 degrees on the sky at the central z=9 plane, yielding arcminute resolution and low sample variance on the scales of interest. 
Along the z-axis, the comoving distance is translated to redshift. The volume thus extends from $z \simeq 4$ to $z \simeq 27$. 
The large redshift extent allows us to capture most reasonable reionization scenarios. 
The densities, velocities, expansion rate, minimum virial mass for star formation, and excursion set barrier (Eq.~\ref{eq:barrier}) are adjusted with redshift. 
The large angular and line-of-sight coverage are important because the ionized regions and large-scale velocity streams that source the kSZ signal appear on scales of tens to hundreds of megaparsecs. 

In addition to the global ionization efficiency parameter, $\zeta$, which sets the timing of reionization, it is important to allow for an additional degree of freedom to capture the influence of physical processes that may impact the {\it duration}, such as feedback effects and recombinations. To do this we add a ``feedback'' parameter, $\alpha$, that makes the effective ionization efficiency an inverse power law of the expected ionization fraction in absence of feedback, $x^*$, which is a measure of the ionizing photons available at a given time\footnote{For a constant ionization efficiency $\zeta$ ($\alpha=0$), the simulation predicts a reionization duration of $\Delta z \sim 3-4$, given the definition of $\Delta z$ below.}: 
\begin{equation} 
\zeta=\zeta_0 \left(\frac{1}{x^*}\right)^\alpha , \quad x^*=\zeta_0 f_{\rm coll}(r=\infty,M_{\rm min}) \, .
\end{equation}

\begin{figure*}.pdf
\bc
\includegraphics[width=18.1cm]{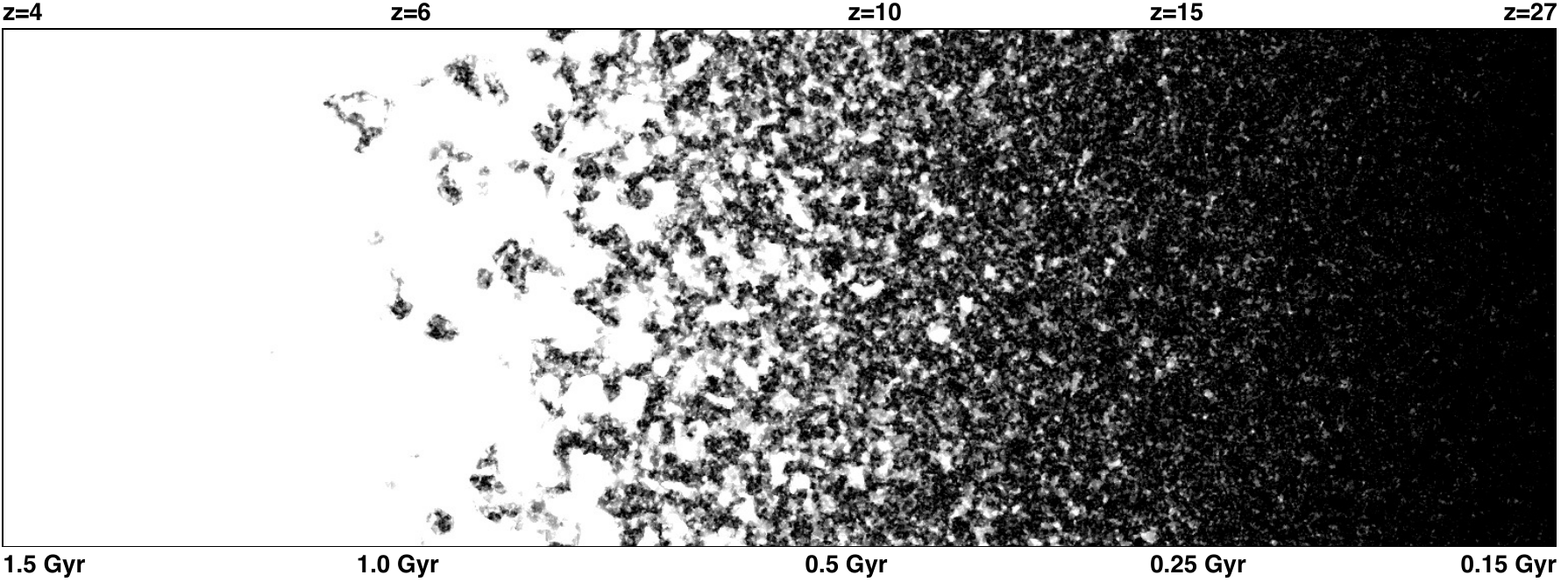} 
\caption{A 20 Mpc/h deep cut through one of our simulation volumes with dimensions 1 $\times$ 3 Gpc/h, corresponding to an angular extent of 8.5 $\times$ 8.5 degrees at the central z=9 plane, and a redshift coverage from $z\simeq 27$ to $z\simeq 4$. Black means fully neutral and white means fully ionized. Note that this is not a result from fitting to data but a random sample from our simulation database.
The ionized fraction in this particular simulation evolves from zero at $z\simeq 20$ to fully ionized at z=5.6.}
\label{fig:xfield}
\ec
\end{figure*}

A database of simulated models is set up for $\zeta \in [10,1000]$ and $\alpha \in [-50, 1]$. 
This leads to $\sim$$\,10000$ simulations that have reionization durations of $\Delta z \in [0.1, 12]$, and ending redshifts of $z_{\rm end} \in [4,16]$\footnote{Note that the Ly-$\alpha$ forest data effectively rule out models that extend past $z_{\rm end}<5$ (see e.g., \citealt{Fan2006,McGreer2011}), and we will impose this as a hard prior when deriving reionization constraints in \S \ref{sec:results}.}. 
Here we have defined $\Delta z=z_{\rm beg}-z_{\rm end}$, $z_{\rm end}=z_{\xhii=0.99}$, and $z_{\rm beg}=z_{\xhii=0.20}$. The choice of a relatively large ionization fraction of $\xhii=0.20$ for $z_{\rm beg}$ is due to the fact that lower ionized fractions are largely made up of ionized regions too small to be probed by the current SPT data, hence we do not attempt to constrain the reionization state preceding this stage.  
An example of a 20 Mpc/h deep projection of an extended reionization simulation with $\zeta=12$, $\alpha=0.8$, and $\Delta z\simeq 9$ is shown in Figure \ref{fig:xfield}.   

For each reionization simulation, we interpolate rays from constant comoving positions to constant angular positions, produce a map of the kinetic SZ with a size of $\simeq 7 \times 7$ degrees, and measure its angular power spectrum. Throughout this paper, power spectra are given in terms of $D_\ell\equiv C_\ell \ell (\ell+1)/2 \pi$ ($\mu K^2$). For each simulation, we also calculate the total integrated optical depth assuming helium is singly ionized alongside hydrogen and doubly ionized at z=3 (the latter assumption has a negligible influence on the results). 

The spatial resolution and box size used lead to convergence of the patchy kSZ amplitude on the relevant angular scale (see the next section). However, the ionized fraction is underestimated during the early stages of reionization, when ionized bubbles smaller than the spatial resolution of the simulation contribute significantly. 
To correctly predict the duration and integrated optical depth we therefore scale the ionized fraction at each fraction to the analytical value $\zeta f_{\rm coll}(r=\infty,M_{\rm min})$, the mean ionization degree of the universe at that redshift.

In these simulations, we assume a  $\Lambda$CDM cosmology with $\Omega_{c} h^2=0.111, \Omega_{b} h^2=0.0222, \Omega_\Lambda=0.736, n_s=0.96, \tau=0.085$, and $\sigma_8=0.8$, consistent with \citet{komatsu11}. 
We discuss the cosmological scaling of the kSZ predictions in \S\ref{sec:homog_ksz} and the effects of uncertainty in the cosmological parameters in \S\ref{sec:kszresults}.

\subsection{The kinetic SZ signal from reionization}

The amplitude and shape of the patchy kSZ power spectrum depend on the redshift evolution of the ionized fraction and  the reionization morphology.  
Here we discuss the dependence of the kSZ power spectrum on the physics of reionization and motivate the template shape used to fit to the SPT data in order to constrain the ionization history. 
 
 The reionization model outlined in \S\ref{sec:reisim} has been shown to agree well with full radiative transfer simulations on scales up to 100 Mpc/h (see e.g., \citealt{Zahn2007,Zahn2010}).
As a generic consequence of the increase in source collapse fraction with time, the size distribution of ionized regions and volume filling factor evolve more slowly at the beginning of reionization than at the end. 
 Also, the roles of feedback effects and recombinations depend on the degree of  ionization. 
If we instead used a model with a parametrized yet physically unmotivated analytic form of $\xhii$, the relation between the redshift corresponding to various ionization fractions $z_{\xhii}$, the kSZ power, and integrated optical depth $\tau$ would be different, potentially biasing constraints on $\xhii$.  

\begin{figure}
\bc
\hspace*{-0.2cm}
\includegraphics[width=8.8cm]{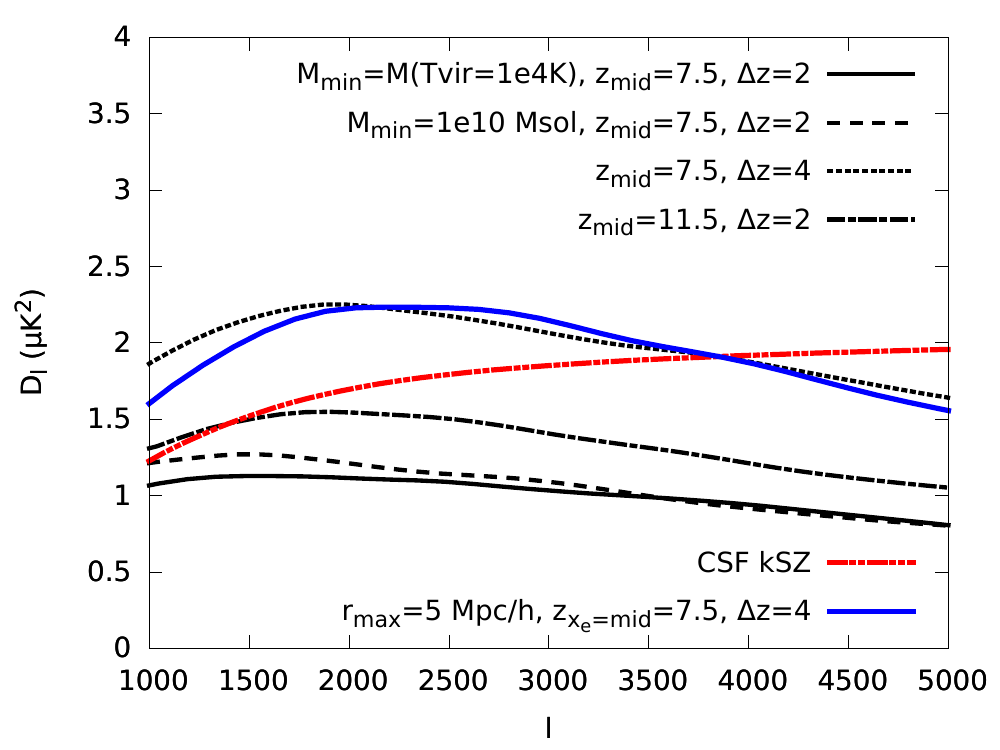}
\caption{kSZ power spectra due to patchy reionization for different $z_{\rm mid}$, defined by $x_e=0.5$ (\textbf{dot-dashed} versus \textbf{dotted black lines}), minimum star forming mass $M_{\rm min}$ (\textbf{dashed} versus \textbf{solid black lines}), and maximum ionized bubble scale (\textbf{blue solid line}). The different $M_{\rm min}$ cases are shown at a smaller $\Delta z$ to allow plotting in the same panel. The \textbf{blue solid} line shows the kSZ power spectrum for a reionization model with $r_{\rm max}=5\,$Mpc/h. The \textbf{red, dot-dashed} line shows our fiducial \hksz\ template (see \S\ref{sec:homog_ksz}).} 
\label{fig:clksz_mmin_zmid_dep}
\ec
\end{figure}

Using radiative transfer simulations, it has been found that the size distribution of ionized regions at fixed $\xhii$ is relatively robust to redshift translations $z_1\rightarrow z_2$ (e.g., \citealt{Zahn2007,McQuinn2007b}). In addition, the angular size corresponding to a given co-moving scale varies little across the redshifts of interest.
The change in the patchy kSZ power spectrum due to a translation from $z_{\rm mid}\simeq 7.5$ to $z_{\rm mid} \simeq 11.5$, where $z_{\rm mid} \equiv z_{\xhii=0.5}$, for similar $\Delta z$ is shown in the black solid and dot-dashed lines in Figure \ref{fig:clksz_mmin_zmid_dep}. 
The shapes of both power spectra are very similar, with the main difference being in their normalization. 
 The amplitude is increased in the $z_{\rm mid}\simeq 11.5$ model due to the higher mean density of the universe at higher redshift, which is counteracted somewhat by linear growth of density and velocity fields toward later times.
  
When moving to very high redshifts, ever rarer sources have to yield enough ionizing photons. 
Such sources are more heavily biased, and their lower number density yields larger and more spherical ionized regions \citep{Zahn2007,McQuinn2007b}. 
The same is  true at low redshifts if feedback effects limit the role of low-mass sources.
For instance, low-mass sources might suffer from thermal feedback onto their host galaxies during early stages of reionization \citep{HultmanKramer:2006dn,Iliev:2006sw}.
The effect on the patchy kSZ power spectrum of increasing the minimum source mass to $10^{10} M_\odot$ (an increase of two orders of magnitude over the fiducial $M_{\rm min}$, discarding $>$\,90\% 
of sources, e.g., \citealt{trac07}) is shown in the black solid and dashed lines of Figure \ref{fig:clksz_mmin_zmid_dep}. 
The impact of source bias on the shape of the patchy kSZ template is found to be small compared to our experimental uncertainty. 

Furthermore, recombinations in dense systems might effectively stall bubble growth early on in reionization, leading to a slowly percolating web of similarly sized ionized regions \citep{FurlanettoOh2005}. To estimate the effect of such a scenario on the kSZ power spectrum, we impose a relatively extreme maximum bubble scale of $r_{\rm max}=5$ Mpc/h. The result is shown in the blue solid line of Figure \ref{fig:clksz_mmin_zmid_dep}. Again, the changes are modest when compared to a model with the same duration but no limit on the bubble scale (dotted curve). 

We conclude that the shape of the kSZ power spectrum due to patchy reionization is relatively robust to the effects of redshift translation, feedback, and recombinations. Most importantly for this work, the amplitude is expected to be proportional to the duration of reionization for uncorrelated ionized regions and fixed $z_{\rm mid}$ (e.g., \citealt{Gruzinov:1998u}).
If the duration doubles in this Poisson process, there are twice as many independently moving ionized regions along the line of sight, which doubles the power. We find this to be a good approximation to the results from our simulations on the angular scales of interest (compare solid versus dotted lines in Figure \ref{fig:clksz_mmin_zmid_dep}).

Given the robustness of the {\it shape} of the patchy kSZ power spectrum to all these changes, we will assume a fixed template shape to derive our reionization results. As base template we choose a model with an efficiency of $\zeta=20$, making this model extend from $z_{\xhii=0.2}\simeq 11.0$ to $z_{\xhii=0.99}\simeq 7.8$, however, as explained above the results are robust to variations in this choice. The base template used to compare to the SPT data was run at 8 times higher spatial resolution than the simulations in the database. The resolution of the latter was chosen to lead to convergence at $\ell=3000$ to allow comparison to the data-derived constraint on the power on that scale, see \S\ref{sec:kszresults}.  

\section{Data}
\label{sec:data}

Here we describe the data used in this work, which serve to constrain either the $\Lambda$CDM  model and primary CMB anisotropy, or the cosmic infrared background (CIB) and the thermal (tSZ) and kinetic (kSZ) Sunyaev-Zel'dovich effects. 
For the $\Lambda$CDM cosmological model, we include measurements of the baryon acoustic oscillation feature from SDSS and 2dFGRS \citep{percival10},  a measurement of the Hubble constant from the Hubble Space Telescope \citep{riess11}, and measurements of the primary CMB from WMAP7  \citep{larson10} and SPT \citep{keisler11}. 
Most importantly for this work, the CMB+BAO+H$_0$ preferred optical depth is  $\tau=0.085\pm 0.014$. 
The optical depth constraint is primarily due to the large-scale polarization and polarization-temperature cross correlation data from WMAP7.

The above datasets do not constrain the Sunyaev-Zel'dovich effects or CIB, which dominate the anisotropies at arcminute scales and SPT wavelengths. 
For these small scale anisotropies, we include new multi-frequency data from the SPT (R11) and the  recent \planckhfi\ CIB bandpowers at 217\,GHz \citep{planck11-6.6_arxiv}. 
The SPT bandpowers are based on observations at 95, 150, and 220\,GHz of $800\,$deg$^2$ of sky. 
From these maps, R11 estimate multi-frequency bandpowers covering angular scales $2000 < \ell \,< 9400$. 
The three-frequency  SPT bandpowers are used to constrain the CIB, tSZ, and kSZ power spectra. 
The \planckhfi\ data aid in constraining the CIB frequency dependence and its angular dependence on large scales.

\section{Model fitting}
\label{sec:fitting}

We use Monte Carlo Markov chain (MCMC) techniques to fit the data with a $\Lambda$CDM cosmological model including lensed primary CMB anisotropy,  tSZ anisotropy, kSZ anisotropy, and foregrounds. 
The model and parameter fitting is described fully in R11. 
Here we briefly describe the full model before focusing on the elements shown by R11 to be important for the kSZ power measurement.  

We adopt the standard, six-parameter, spatially flat,  
$\Lambda$CDM cosmological model to predict the primary CMB temperature anisotropy. 
R11 consider model extensions including running of the spectral index, massive neutrinos, and freedom in the number of neutrino species, finding no effect on the kSZ constraints. 
R11 also show that the systematic uncertainty in kSZ power due to uncertainty in the CMB lensing potential is much smaller than the derived kSZ uncertainties. 

We include seven free parameters for the small-scale temperature anisotropies from the tSZ and kSZ effects, and the CIB. 
We also include radio galaxies and galactic cirrus with strong priors as described by R11.  
Two parameters are the amplitudes of the tSZ and kSZ power spectra; the remaining five parameters describe the CIB model. 
The chosen tSZ template is the baseline model from  \citet{shaw10}. 
R11 show that the resulting kSZ power spectrum results are insensitive to the shape of the tSZ template. 
The kSZ template and CIB model are critical for the reionization constraints in this work, and will be described next. 

\subsection{Kinetic Sunyaev-Zel'dovich effect}

We adopt two main templates for the kSZ power spectrum. 
The first template includes only the contribution from patchy reionization (see \S\ref{sec:pkszmodeling}). 
The second and more realistic template includes roughly equal contributions from the reionization and  post-reionization epochs. 
We henceforth refer to the post-reionization, homogeneously ionized, component as the
``homogeneous kSZ'' signal, and that from reionization as the ``patchy kSZ''  signal. 
We discuss the \hksz\ model next. 

\subsubsection{Homogeneous kinetic Sunyaev-Zel'dovich effect}
\label{sec:homog_ksz}

For the \hksz\ signal, we adopt the cooling plus star-formation (CSF)
model presented by \citet{shaw11}. 
This model is constructed by
calibrating an analytic model for the \hksz\ power spectrum with a hydrodynamical simulation including metallicity-dependent radiative cooling and star-formation. 
\citet{shaw11} measured the power spectrum of gas density fluctuations in the simulation over a range of redshifts,
and used this to calculate the \hksz\ power spectrum. 
The approximate scaling for this model with cosmological parameters is given  by
\begin{eqnarray}
\Dhksz  &\simeq& 1.9\,\mu{\rm K}^2 \left(\frac{h}{0.71}\right)^{1.7} \left(\frac{\sigma_8}{0.80}\right)^{4.5} \left(\frac{\Omega_b}{0.044}\right)^{2.1}\times\nonumber\\
&& \left(\frac{\Omega_m}{0.264}\right)^{-0.44} \left(\frac{n_s}{0.96}\right)^{-0.19}.
\label{eq:ksz_scaling}
\end{eqnarray}
This scaling (but not amplitude) is also a good approximation to the cosmological dependence of the patchy kSZ component. We vary the predicted post-reionization \Dhksz\ with $z_{\rm end}$ when interpreting the results in \S\ref{sec:results}. The CSF model kSZ component for $z_{\rm end}=8$ is shown in the red line in Figure \ref{fig:clksz_mmin_zmid_dep}. 
The model utilized here assumes helium is singly ionized between $3 \leq z \leq z_{\rm end}$ and doubly ionized for $z < 3$.

\subsection{Dusty galaxies (CIB)}

The CIB is produced by thermal emission
from dusty star-forming galaxies (DSFGs) over a very broad range in
redshift \citep{lagache05,marsden09}.  The dust grains, ranging in
size from a few molecules to 0.1 mm, absorb light at wavelengths
smaller than their size, and re-radiate it at longer wavelengths.
Sufficient absorption occurs to account for roughly equal amounts of
energy in the CIB and in the unprocessed starlight that makes up the
optical/UV background \citep{dwek98,fixsen98}.

The power spectrum of these DSFGs will have Poisson and clustered components.  
Both CIB components have more power than the kSZ power spectrum on the angular scales and photon frequencies of interest (except at 95\,GHz where the powers are comparable). 
The Poisson component shows up as ${\it D}_\ell \propto \ell^{2}$ since the galaxies are small (effectively point sources) compared to the angular scales  probed here. 
We model the clustered component by a power-law of the form ${\it D}_\ell \propto \ell^{0.8}$, a shape motivated by recent observations (\citealt{addison11}, R11). 
We have explored allowing freedom in the angular shape via freeing the power-law exponent or adding a linear-theory template, and find the additional shape parameter increases the kSZ uncertainties by $\sim$10\% without changing the mean.

The frequency dependence of the CIB is a key question for the kSZ measurement. 
Essentially, the CIB is well measured at $\ge$\,220\,GHz by SPT, Planck, and other experiments, but the signal-to-noise falls off towards longer wavelengths. 
The CIB power at 95\,GHz is  an extrapolation based on a fit to the modeled frequency dependence. 
Following previous work, we adopt the phenomenological modified black-body model:
\begin{equation}
\label{eqn:cib}
\eta_{\nu}  = \nu^\beta B_\nu(T).
\end{equation}
$\nu$ is the observing frequency, and $B_\nu(T)$ is the black-body spectrum for temperature T. 
T and $\beta$ are free parameters with uniform priors $T \in [ 5,35 \,K]$ and $\beta \in [0,2]$. 
R11 also examined the single-SED model presented by \citet{hall10}.
The resulting kSZ power upper limits were similar when allowing for the possibility of correlation between the CIB and tSZ and 20\% tighter when any correlation was assumed to be negligibly small.

We assume the clustering and Poisson components have the same frequency scaling in this work. 
Implicitly this is equivalent to assuming that the source populations and redshift distributions are similar. 
\citet{hall10} argued that any difference in the frequency scaling should be small based on CIB simulations. 
As mentioned in R11, the data have some preference for differing spectral indices when considering the 95 - 220\,GHz alone, however this preference vanishes when \planckhfi\ 353\,GHz data is added.

\subsection{Correlation between the tSZ and CIB}
\label{sec:cibtsz}

An important component of our model is the potential for a spatial
correlation between the DSFGs that produce the
CIB and the groups and clusters that generate the tSZ power spectrum.
At $\ell = 3000$, most of the tSZ power is predicted to originate from
clusters of mass greater than $ 5\times 10^{13} \hmsun$ and redshift
less than 1.5 \citep{shaw09, battaglia11, trac11}.  If the galaxies
residing in these halos contribute a significant fraction of the CIB
power at the SPT frequencies, then the spatial correlation between the
tSZ and CIB signal will be non-negligible.  Below 220 GHz, the tSZ
effect produces temperature decrements; we thus expect an
anti-correlation between the CIB and tSZ fluctuations. Note that
cross-spectra involving the 220 GHz band -- the tSZ null frequency --
may still contain power due to tSZ-CIB correlation as the CIB signal couples with
the tSZ at the second frequency.

Following R11, we model the contribution of a tSZ-CIB correlation to the measured power spectrum as:
\small
\begin{equation}
\label{eqn:tszcib}
D^{\rm tSZ-CIB}_{\ell,\nu_i \nu_j} = \xi_\ell \left[ \sqrt(D^{\rm tSZ}_{\ell,\nu_i \nu_i}  D^{\rm CIB}_{\ell,\nu_j \nu_j} )  +   \sqrt(D^{\rm tSZ}_{\ell,\nu_j \nu_j}  D^{\rm CIB}_{\ell,\nu_i \nu_i})   \right] \;.
\end{equation}
\normalsize 
\noindent Here $D^{\rm tSZ-CIB}_{\ell,\nu_i \nu_j}$ is the power due
to correlations, $D^{tSZ}_{\ell,\nu_i\nu_i}$ is the tSZ power
spectrum at frequency $i$, and $D^{CIB}_{\ell,\nu_i\nu_i}$ is the sum
of the Poisson and clustered CIB components.  $\xi$ is the correlation
coefficient; we define $\xi_{3000}$ as the amplitude of this
correlation at $\ell = 3000$.  We consider two templates for the
spatial dependence of the correlation, one flat and one rising with
$\ell$.  The reasoning behind these templates is discussed later in this section.

R11 measure the magnitude of the tSZ-CIB correlation by introducing
$\xi$ as a free parameter in their model fit to the SPT band powers,
assuming an $\ell$-independent template for $D^{\rm
  tSZ-CIB}_{\ell}$. They find $\xi_{3000} = -0.18 \pm 0.12$,
consistent with a fraction of the CIB emission being spatially
coincident with the tSZ signal.  However, \citet{george11} extrapolate
the 24\,$\mu$m flux of galaxies in their (X-ray selected) group
catalog to 2\,mm and find a very small signal.  Even allowing for
uncertainties in this extrapolation, they conclude that the
contamination of DSFGs to the tSZ signal is no more than a few
percent.  This would imply only a very weak spatial correlation
between these signals.  A robust measurement of the total DSFG
emission in groups and clusters at mm/sub-mm wavelengths would place
tight observational constraints on $\xi_\ell$.

There currently exist no published theoretical estimates of the
magnitude of the tSZ-CIB correlation. However, we find that the
publicly available simulations of \citet{sehgal10} predict a large
anti-correlation.  Taking the cross-spectrum of their simulated
CIB-only and tSZ-only maps at 148\,GHz, we find $\xi_{3000} = -0.37$.

We make new predictions for the tSZ-CIB correlation by combining the
tSZ model of \citet{shaw10} with the halo model based CIB calculations
of \citet{shang11}. In brief, the \citet{shang11} model populates
halos and sub-halos with DSFGs utilizing the halo mass function of
\citet{tinker08} and the sub-halo mass function of \cite{wetzel10}. It
is assumed that the relation between galaxy luminosity and sub-halo
mass is lognormal, with free parameters for the overall normalization,
characteristic mass, and redshift evolution. These parameters are
calibrated by comparing the model predictions with the recent {\it
  Planck} measurements of the CIB power spectrum
\citep{planck11-6.6_arxiv}. \citet{shang11} explore model parameter
constraints for various scenarios in which the characteristic dust
temperature of DSFGs, the redshift evolution of the subhalo mass -
DSFG luminosity relation, or the shape of their SED are alternatively
fixed or allowed to vary.  For each of these scenarios \citep[labeled
  ``cases 0 to 5'' in][]{shang11}, we calculate the correlation
coefficient, $\xi$.

We find that the predicted value of $\xi_{3000}$ varies between $-0.02
> \xi_{3000} > -0.34$, well within the range found by R11. $\xi$
depends sensitively on the assumed redshift evolution of DSFG
luminosity (or equivalently, the evolution of their star-formation
rate), which always increases towards higher redshift. For models
in which the luminosity evolves rapidly, a larger fraction of the CIB
power is contributed by high-redshift ($z>2$) objects and thus the
magnitude of $\xi$ decreases.

\begin{figure}[t]
\centering
\includegraphics[width=8cm]{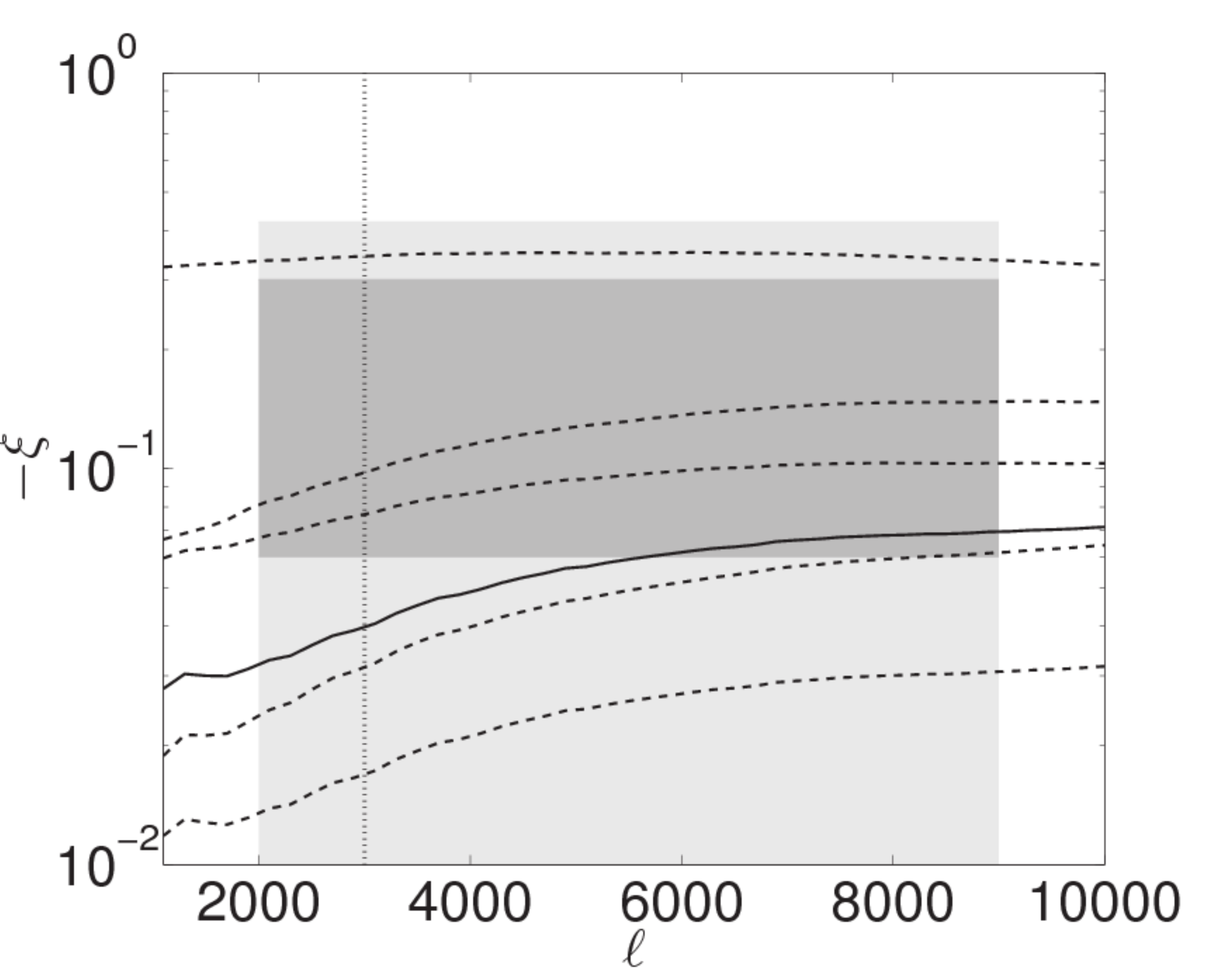}
\caption{The correlation coefficient $\xi$ calculated for the various
  CIB model cases presented in \citet{shang11}. 
  The correlation coefficients at $\ell
  = 3000$ are (from top to bottom) $\{-0.34, -0.09, -0.07, -0.04,
  -0.03, -0.02\}$. The slope of the correlation with $\ell$ becomes
  steeper as $|\xi|$ decreases. The magnitude of $\xi$ is most sensitive to the redshift evolution of the halo mass - DSFG luminosity relation; a more rapid increase in luminosity with redshift shifts the peak of the CIB to higher redshifts and therefore reduces the strength of the tSZ-CIB correlation. The shaded regions represent the 1 and 2~$\sigma$ constraints on $\xi$  presented in R11 based on an $\ell$-independent correlation template, over the range of angular scales probed by SPT. In this work we also explore the effect, on the derived kSZ power, of using an $\ell$-dependent template with a free amplitude, shown in the solid line.}
\label{fig:tsz_cib_templates}
\end{figure}

The $\ell$-dependence of $\xi$ also varies between our CIB models. In
Figure \ref{fig:tsz_cib_templates}, we plot $\xi_\ell$ for each of the
model cases presented by \citet{shang11}. The values of $\xi_{3000}$
are given in the caption. The shaded region represents the 1- and
2~$\sigma$ constraints on $\xi$ obtained by R11. We find a general
trend that the correlation coefficient is a stronger function of
$\ell$ for models in which the overall amplitude is closer to
zero. This is straightforward to understand; models with a lower value
of $|\xi|$ are those for which high redshift galaxies contribute a
larger fraction of the total CIB power.  Most of the tSZ-CIB
correlation then comes from higher redshift halos which principally
contribute to the tSZ power at small scales. Therefore, the
correlation coefficient increases with $\ell$. 

We will explore the dependence of constraints on the kinetic SZ power spectrum on the shape of $\xi_\ell$, focusing on the model shown in the solid curve, which has an average correlation coefficient of 0.04, but letting the amplitude freely vary. Note that this is conservative in that models with steeper amplitude have correlations close to zero, however we will not impose a prior.  

\section{Results for the Optical Depth and patchy kSZ power spectrum} \label{sec:tauksz}

In this section, we briefly review the optical depth measurement from WMAP7 in the context of reionization. 
We then present the SPT patchy kSZ power spectrum measurement, and discuss factors that can affect this result. 

\subsection{Optical Depth}
\label{sec:tauresults}

\begin{figure*}[ht!]
\resizebox{\hsize}{!}{
\plottwo{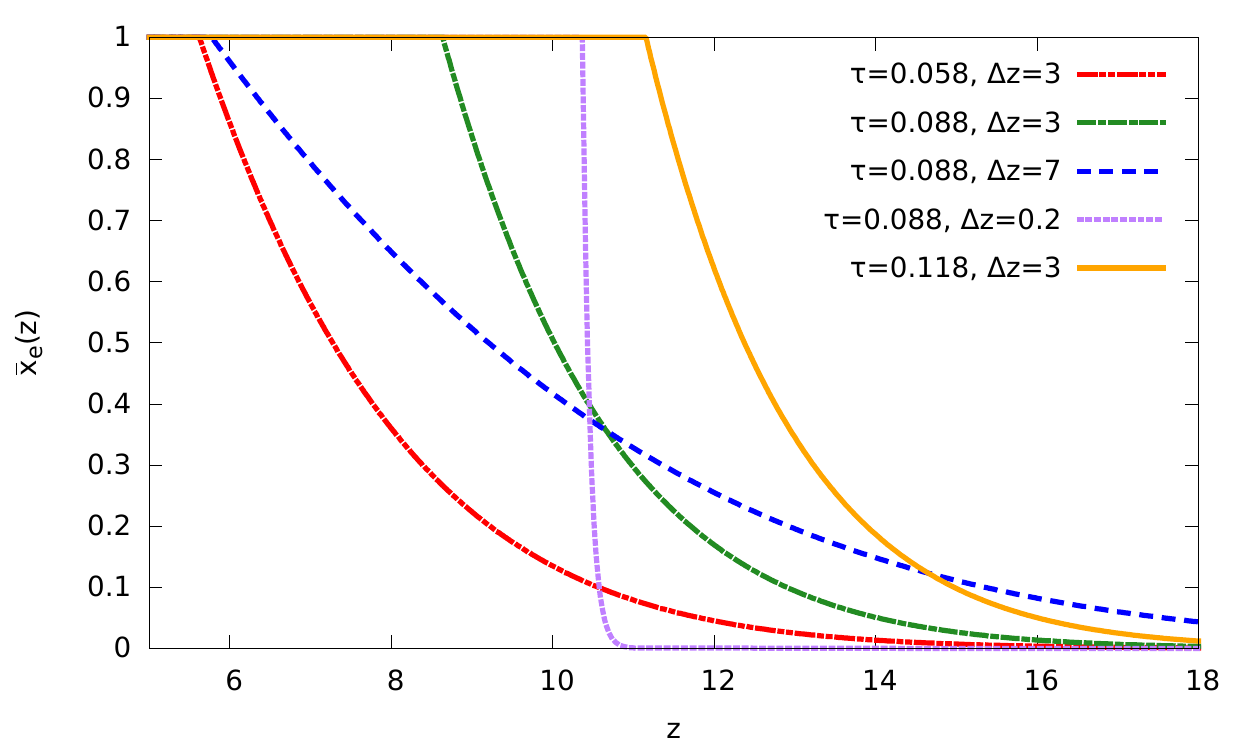}{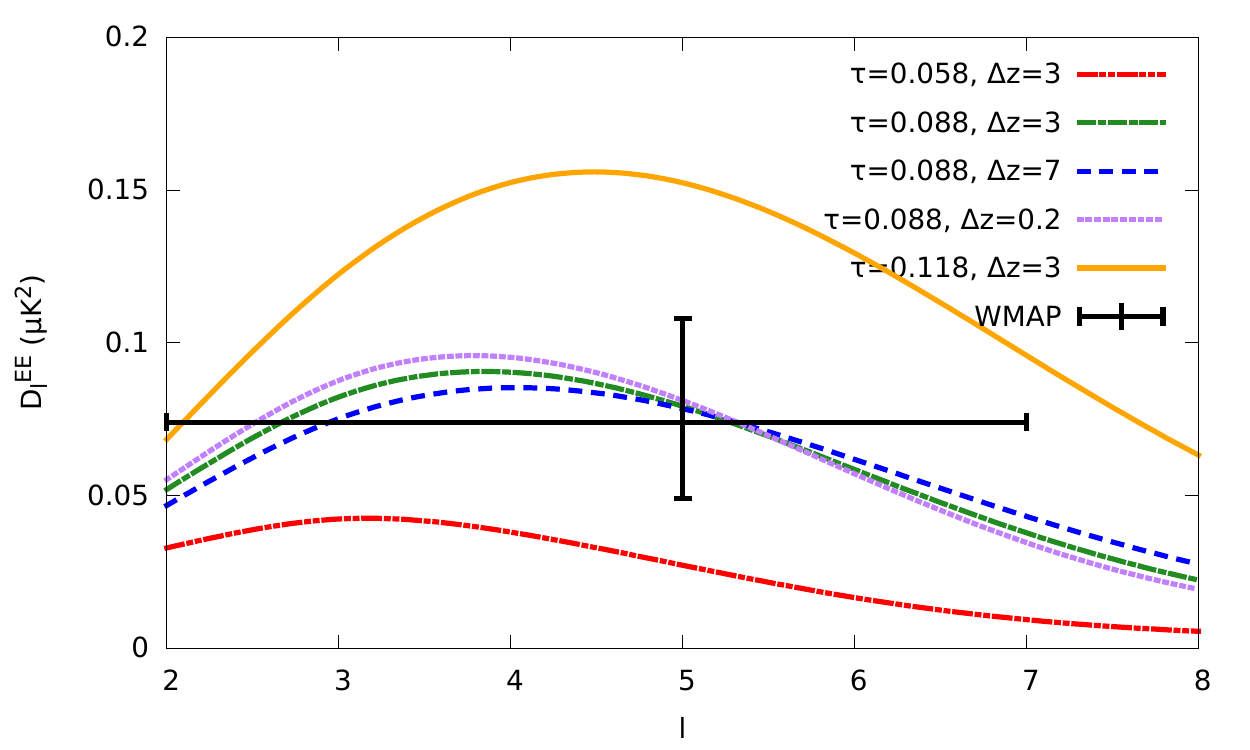}}
\caption{Ionization history and integrated optical depth for different reionization models. 
The color-coding of model lines is the same between panels. 
\textbf{\textit{Left panel:}}  Evolution of the ionized fraction for each model. 
\textbf{\textit{Right panel:}}  EE power spectra for the models, together with WMAP7 \citep{larson10} band power constraint (black data point). 
The \wmap\ measurement of E-mode power in the band $2<\ell <7$ does not restrict the duration of reionization.}
\label{fig:qzwmaptau}
\end{figure*}

\wmap\ measures the optical depth to reionization to be $\tau=0.088\pm 0.014$ \citep{larson10}. 
Figure \ref{fig:qzwmaptau} provides intuition on how this constrains the epoch of reionization. 
In the left panel, we show ionization histories for several models from our simulation database. 
The right panel plots the large-scale polarization signal as calculated by CAMB \citep{lewis99} from the simulated ionization histories along with the WMAP7 bandpowers. 
The near instantaneous model at $z_{\rm mid} = 10.5$ shown in the purple dotted curve, as well as the more extended green dot-dashed curve (${\Delta} z \simeq 3$) and very extended blue dashed curve (${\Delta} z \simeq 8$) are good fits to the \wmap\ data, as shown in the right hand panel of Figure \ref{fig:qzwmaptau}. The large scale polarization data do not allow one to discern between shorter and longer duration models. 
However, the red dot-dot-dashed and orange solid curves with $z_{\rm mid} \simeq $ 7 and 12.5 are poor fits to the \wmap\ data.  
For the data used in this paper (CMB+BAO+H$_0$), the optical depth constraint  is $\tau=0.085\pm 0.014$. 
This corresponds to a $\sim 0.25$ shift to lower redshifts compared to using the WMAP-only optical depth constraint; we use $\tau=0.085\pm 0.014$ for all reionization constraints in this work.

\subsection{kSZ power}
\label{sec:kszresults}

\begin{figure}[t]
\centering
\hspace*{-0.5cm}
\includegraphics[width=9.5cm]{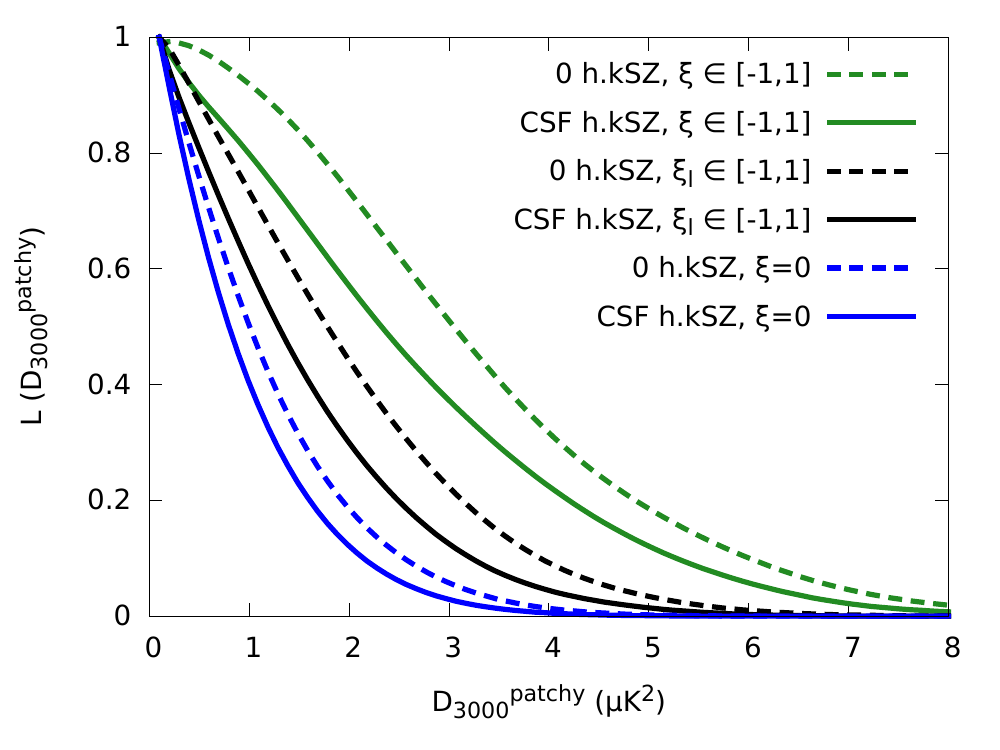}
\caption{Constraints on the amplitude at $\ell = 3000$ of patchy kSZ fluctuations from the SPT data for different \hksz\ and tSZ-CIB correlation models. The upper limits tighten with the addition of \hksz\  power and with more restrictive assumptions about the tSZ-CIB correlation.}
\label{fig:clconstraints}
\end{figure}

The SPT constraints on the kSZ power spectrum are summarized in Figure \ref{fig:clconstraints} and Table \ref{tab:pksz}.
The table contains 95\% confidence upper limits on the patchy kSZ power at $\ell = 3000$ for different \hksz\ and CIB models.

The kSZ contributions from various redshifts have the same frequency dependence, and only small differences in angular dependence (as shown in Figure \ref{fig:clksz_mmin_zmid_dep}). We find that the 2008 and 2009 SPT data presented by R11 do not yet  significantly discriminate between the patchy and \hksz\ template shapes. 
As a result, the modeled \hksz\  power affects the level of patchy kSZ allowed by the data. With more \hksz\ power,  less patchy kSZ power is allowed. 
We note, however, that the patchy kSZ likelihood curves in Figure \ref{fig:clconstraints} shift by relatively little when changing the \hksz\ model by $\sim 2\, \mu K^2$.  
This is because the likelihood curves peak at or below  zero kSZ power ($-1.2\,\sigma$ without tSZ-CIB correlation), and a positivity prior is applied. As will be discussed below, the SPT data prefer a non-negligible tSZ-CIB anti-correlation together with an increased kSZ amplitude.

We consider two models for the \hksz\ signal. 
The first, zero \hksz\ (dashed lines in Figure \ref{fig:clconstraints}), is the most conservative when determining upper limits on the patchy kSZ signal. 
However, it is physically unmotivated. 
The second case (solid lines) is the CSF model described in \S\ref{sec:homog_ksz} which is 
our best estimate for the \hksz\ signal. 
We scale the \hksz\ signal according to the cosmological parameters at each step in the chain following Eqn.~\ref{eq:ksz_scaling}. 
We tabulate  the patchy kSZ upper limits with these two \hksz\ models in the first and second rows of Table \ref{tab:pksz}. 

To constrain the epoch of reionization, we use kSZ power constraints based on a fixed template set to the sum of the CSF and patchy kSZ models in the fiducial cosmology. 
Constraints on the total kSZ power with this template are found in the third row of Table \ref{tab:pksz}. 
The small error caused by our choice of a fixed ratio between patchy and \hksz\  templates amounts to, at most, a fraction of the observed 15\% difference between the kSZ upper limits for the CSF and patchy templates.


A significant uncertainty in the current kSZ constraints is related to the unknown tSZ-CIB correlation.  
As discussed in  \S\ref{sec:cibtsz},  the amount of correlation is highly uncertain; we therefore explore three cases for the correlation. 
For no correlations, shown in the blue lines of Figure \ref{fig:clconstraints} and left column of Table  \ref{tab:pksz}, the inferred 95\% confidence upper limits on the patchy kSZ power at $\ell=3000$ are 2.5 and $2.1\,\mu{\rm K}^2$ with the zero and CSF \hksz\ models respectively. 
Our second patchy kSZ constraint (black lines in Figure \ref{fig:clconstraints} and center column of Table  \ref{tab:pksz}) uses a rising tSZ-CIB correlation shape based on the modeling in \S\ref{sec:cibtsz}. 
This correlation shape is the solid line in Figure \ref{fig:tsz_cib_templates}, and corresponds to the $\xi_{3000}=-0.04$  Shang model case. 
However, the amplitude of the tSZ-CIB correlation is allowed to vary freely in the MCMC. 
The third case has an $\ell$-independent tSZ-CIB correlation of unknown amplitude. 
The data prefer a larger anti-correlation in this final case; the 95\% confidence range is $\xi_{3000} \in [-0.28, 0.03]$ for the rising correlation shape and $\xi_{3000} \in [-0.43, -0.01]$ for the  $\ell$-independent correlation shape.
As a result, the third case  yields the most conservative constraints on the kSZ effect. 
The 95\%  confidence upper limits on the patchy kSZ power weaken to  5.5 and $4.9\,\mu{\rm K}^2$ with the zero and CSF \hksz\ models respectively. 

\begin{table}
\bc
\caption{kSZ Upper Limits}
\begin{tabular}{c|ccc}
\hline
\hline
 kSZ power &  $\xi=0$ & $\xi_\ell\in[-1,1]$& $\xi\in[-1,1]$\\
  \hline
$D^{\rm patchy}_{3000}$ (0 h.kSZ) & 2.5 $\mu {\rm K}^2$&  3.7 $\mu {\rm K}^2$&  5.5 $\mu {\rm K}^2$\\
$D^{\rm patchy}_{3000}$ (CSF h.kSZ) &  2.1 $\mu {\rm K}^2$& 3.1 $\mu {\rm K}^2$&  4.9 $\mu {\rm K}^2$\\
\\
\hline
\\
$D^{\rm kSZ}_{3000}$  &  2.7 $\mu {\rm K}^2$& 4.1 $\mu {\rm K}^2$&6.1 $\mu {\rm K}^2$\\
\end{tabular}
\label{tab:pksz}
\ec
\tablecomments{95\% confidence upper limits on the kSZ power at $\ell = 3000$ for different assumptions about the \hksz\ model and tSZ-CIB correlation, $\xi$.
The first and second rows have limits on the patchy kSZ power assuming either zero \hksz\ power or the CSF \hksz\ model. 
The third row shows constraints on the total kSZ power for a combined patchy + CSF \hksz\ template. 
The combined template is used to infer constraints on the epoch of reionization. 
The first column assumes the most restrictive model for the correlations, $\xi = 0$, and therefore leads to the tightest kSZ limits.  
The second column allows free correlations with the rising model shape (see \S\ref{sec:cibtsz} and the solid line in Figure \ref{fig:tsz_cib_templates}) while the third column allows a free, $\ell$-independent tSZ-CIB correlation.
The latter yields the weakest kSZ upper limits.}
\end{table}

\begin{figure}[t]
\centering
\hspace*{-0.5cm}
\includegraphics[width=9cm]{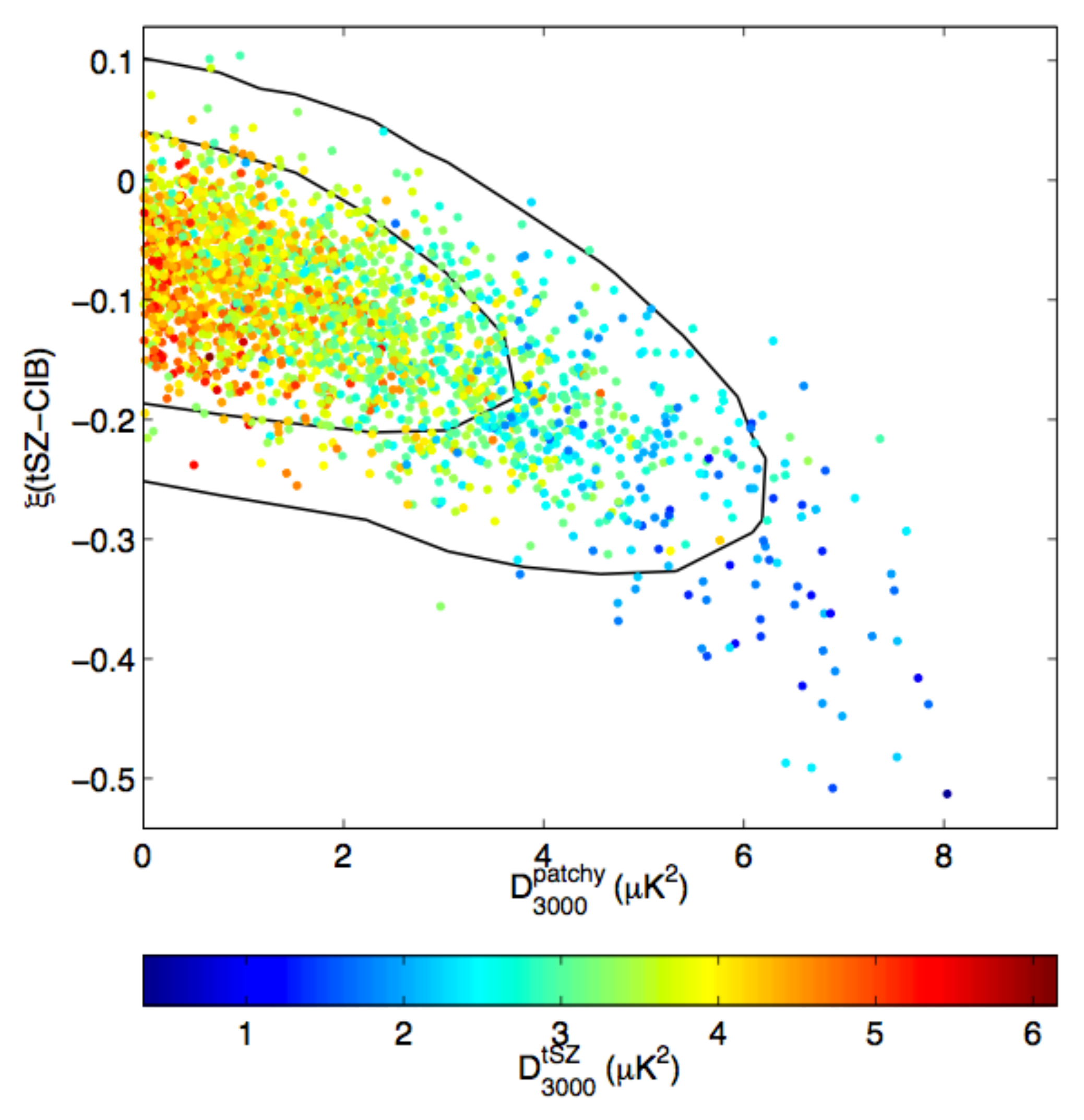} 
\caption{Degeneracy between the patchy kSZ and tSZ-CIB correlation fit to the SPT data. The dots are de-correlated Markov chain samples, color coded by the tSZ  power. The inner/outer contours are 68\% and 95\% regions, respectively. There is little room for patchy kSZ power with $\xi=0$. Allowing $\xi$ to be free, the patchy kSZ contribution can be large at the cost of small values of  tSZ power.}
\label{fig:kszgamma}
\end{figure}

Figure \ref{fig:kszgamma} shows the degeneracy between the patchy kSZ contribution, and the tSZ-CIB correlation, $\xi$. 
The figure assumes no homogeneous kSZ contribution and an $\ell$-independent tSZ-CIB correlation; qualitatively the result is independent of these assumptions.  
It is evident that the kSZ power increases with
the anti-correlation (more negative values of the correlation, $\xi$).
Anti-correlation has the opposite effect on the tSZ power.  The
decrease in tSZ power with increasing anti-correlation is shown in the
color-coding of the independent samples, ranging from values of $4\,
\mu K^2$ for the tSZ with zero tSZ-CIB correlation, to small values of $2\,\mu K^2$ for a -0.5 correlation.  

\begin{figure*}[t]
\centering
\includegraphics[width=18cm]{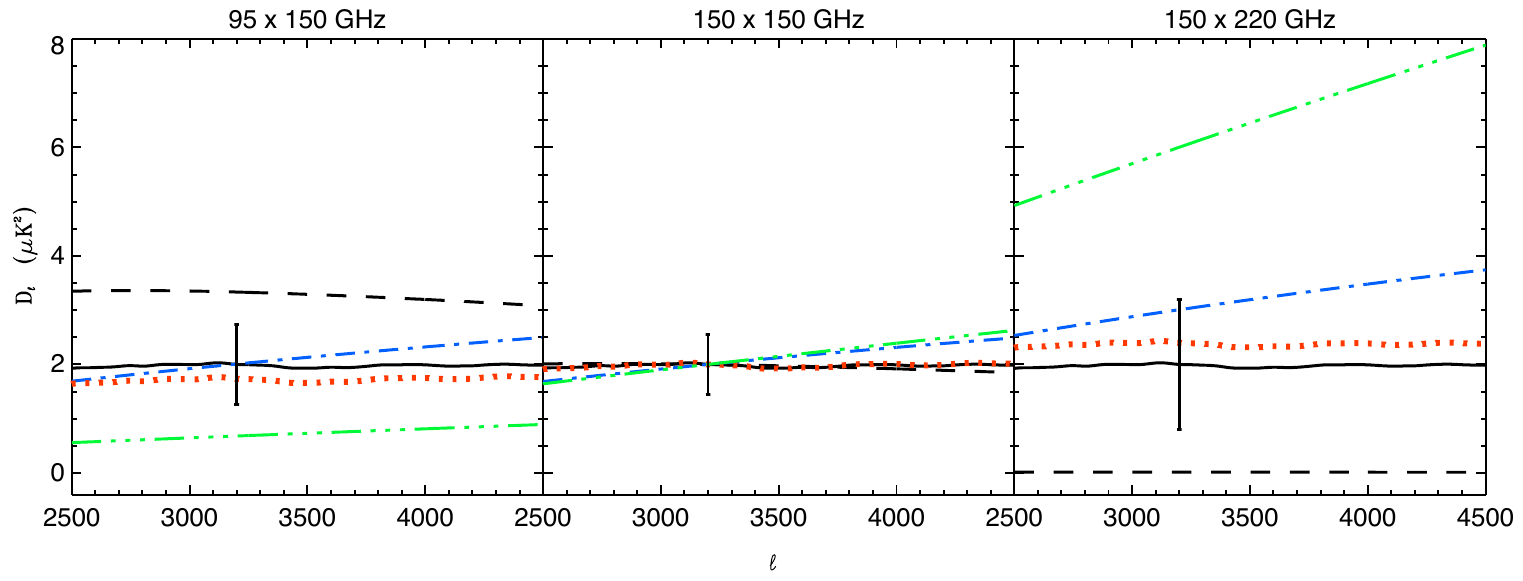} 
\caption{
Key model components in the three most important SPT frequency combinations. 
All model components have been normalized to equal $D_\ell=2 \,\mu {\rm K}^2$ at $\ell=3200$ in the 150x150 GHz spectrum. 
From left to right, the panels show the power spectra for the $95\times 150\,$GHz, $150\times 150\,$GHz, $150\times 220\,$GHz bands. 
The SPT uncertainty on the average power across the multipole range $\ell\in[2600,4600]$ is displayed by the black error bars (also centered at $D_\ell=2 \,\mu {\rm K}^2$ at $\ell=3200$). 
The tSZ-CIB cross power (\textbf{blue dot-dashed line}) increases slightly with frequency, as does a linear combination of tSZ and kSZ (\textbf{red dotted line}). 
As the tSZ (\textbf{black dashed line}) has the opposite frequency dependence, a tSZ-CIB anti-correlation allows for a \textbf{larger} kSZ amplitude (shown as the combined patchy+homogeneous template in the \textbf{black solid line}). The clustered component of the CIB is shown as well (\textbf{green dot-dashed line}); it has a much steeper frequency scaling.}
\label{fig:multifreq}
\end{figure*}

Decreasing tSZ
power with increasing anti-correlation contradicts our naive
expectations.  In single-frequency band powers (e.g., only 150\,GHz),
increasing tSZ-CIB anti-correlation would increase the allowed tSZ
contribution for any observed power level. The anti-correlation term
is negative in this band, thereby allowing more tSZ power. 
However,
this picture changes when we consider the SPT multi-frequency data.  
In Figure \ref{fig:multifreq}, we show  relevant model components across the three deepest SPT frequency combinations ($95\times150\,$GHz, $150\times150\,$GHz, and $150\times 220\,$GHz from left to right). 
We have normalized each model component to 2\,$\mu{\rm K}^2$ at 150\,GHz and $\ell=3000$. 
Note that a tSZ-CIB anti-correlation would lead to a negative power rather than the positive signal (shown in the blue dot-dashed line) in the plot.  
As shown, a tSZ-CIB anti-correlation reduces the power at
$150\times150$ GHz (which could be compensated by increasing tSZ power, shown in the black dashed line)
but it also has $\sim$50\% larger effect at $150\times220$ GHz (where
there is effectively zero tSZ power).  Conversely, a tSZ-CIB
anti-correlation reduces the power at $95\times150$ GHz by a similar
amount to $150\times150$ GHz, whereas there is more tSZ power in the
95\,GHz band.  Adding kSZ power (shown as the {\it combined} template in the black solid line) more effectively compensates for the tSZ-CIB
correlation term than adding tSZ power in the most sensitive
combinations of the three frequencies. A combination of the kSZ and tSZ does even better, as qualitatively shown by the red dotted line for $1.2 D^{\rm kSZ}_{\ell} - 0.2 D^{\rm tSZ}_{\ell}$. 

\section{Constraints on the Epoch of Reionization} \label{sec:results}

In this section, we present constraints on the duration, beginning,  and end of reionization. 
We first interpret the kSZ power in terms of the duration of reionization. 
We then combine the kSZ constraint from SPT with the integrated optical depth constraint from  WMAP7 to constrain the evolution of the ionized fraction as function of redshift, $\xhii$. 
Finally, we use the measurement of $\xhii$ to present limits on $z_{\rm end}$ and $z_{\rm beg}$.

We derive constraints on the history of reionization by comparing the experimental constraints on the total optical depth and the  kSZ power to the predicted values for each model in our simulation database. 
Recall that we use the reionization simulations to predict the patchy kSZ contribution. 
We then add this patchy power to the power predicted by the CSF \hksz\ model at the fiducial cosmology for the $z_{\rm end}$ of this simulation to estimate the total kSZ power. 
Uncertainties in the cosmological parameters within the Monte Carlo Markov chains used in this paper propagate into a 12.5\% uncertainty in the kSZ power spectrum amplitude according to Eq.~\ref{eq:ksz_scaling}. 
We therefore include a 12.5\% Gaussian scatter in $D^{\rm kSZ}_{3000}$ when deriving reionization constraints. 
This is a good approximation as the kSZ power constraint from the current data is independent of the CMB-derived cosmological parameters. 

We marginalize over models in our simulation database, weighting each model by the joint \wmap+SPT likelihood. We use a conservative prior of $\xhii=1$ at $z\leq 5$ based on the model-independent constraints of \citep{Fan2006,McGreer2011}.  

\subsection{Constraints on the duration of the epoch of reionization, $\Delta z_{\rm rei}$}
\label{sec:dzresults}

Here we present constraints on the duration of reionization, defined as $\Delta z_{\rm rei}\equiv z_{\xhii=0.20}-z_{\xhii=0.99}$.
These constraints are primarily due to the SPT kSZ measurement; \wmap\ data alone do not discriminate between reionization models of varying durations. However, by constraining the integrated optical depth, the \wmap\ data indirectly affect the duration inferred from the SPT-derived kSZ constraint since models with given duration have a larger signal at higher redshift due to the higher mean density at higher redshift (see Figure \ref{fig:clksz_mmin_zmid_dep}). 

\begin{figure}[t]
\centering
\hspace*{-.25cm}
\includegraphics[width=9cm]{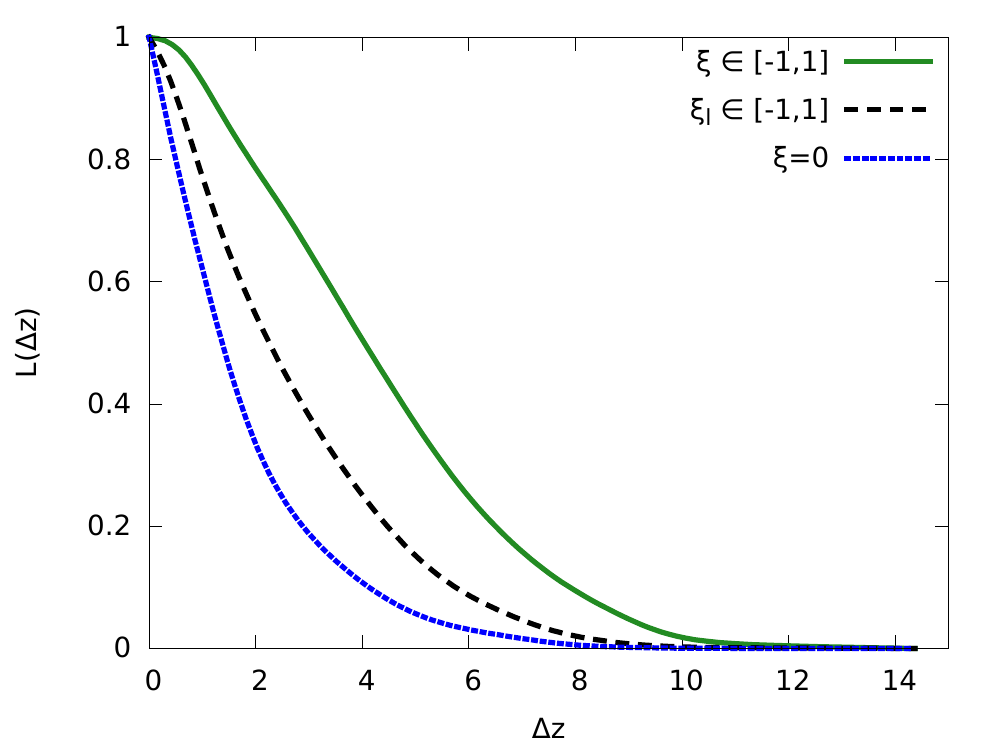}
\caption{Constraints on the duration of reionization for three assumptions about the tSZ-CIB correlation; free amplitude and scale-independent correlation (\textbf{solid green line}), free amplitude and rising correlation with amplitude consistent with modeling of low correlation simulations (\textbf{dashed black line}); and with correlation set to zero (\textbf{dotted blue line}). 
We have assumed the CSF \hksz\ model for the post-reionization kSZ signal throughout. The \wmap\ large-scale polarization data enter by constraining the total electron scattering optical depth but are not able to constrain the duration of reionization. }
\label{fig:ldeltaz}
\end{figure}

In Figure \ref{fig:ldeltaz}, we show constraints on $\Delta z$ for the three tSZ-CIB correlation cases described in \S\ref{sec:kszresults}. 
Under the assumption of no tSZ-CIB correlation, we find a 95\%
upper limit on the duration of $\Delta z < 4.4$. 
The rising correlation shape template leads to an intermediate limit of $\Delta z < 6.2$.
With the $\ell$-independent correlation, we find the weakest 95\% upper limit of $\Delta z < 7.9$. Note from the left panel of Figure \ref{fig:qzwmaptau} that a shorter redshift interval follows the midpoint of the epoch of reionization than precedes it. As a result, only a fraction of the duration in Figure \ref{fig:ldeltaz} can lie at lower redshift than the midpoint $z\sim 10.25$.
 
\subsection{Constraints on the evolution of the ionized fraction $\xhii$}
\label{sec:xzresults}

In this section, we show that we can use the CMB data to constrain the history of reionization, $\xhii$, by combining the constraints on the integrated optical depth and duration of the epoch. To this end, we compute the posterior distributions of $z(\xhii)$ for all values of $\xhii$ using our simulation grid given the constraints on $\tau$ and the amplitude of the kSZ power spectrum, $D_{3000}^{\rm patchy}$. 

\begin{figure}
\centering
\includegraphics[width=8cm]{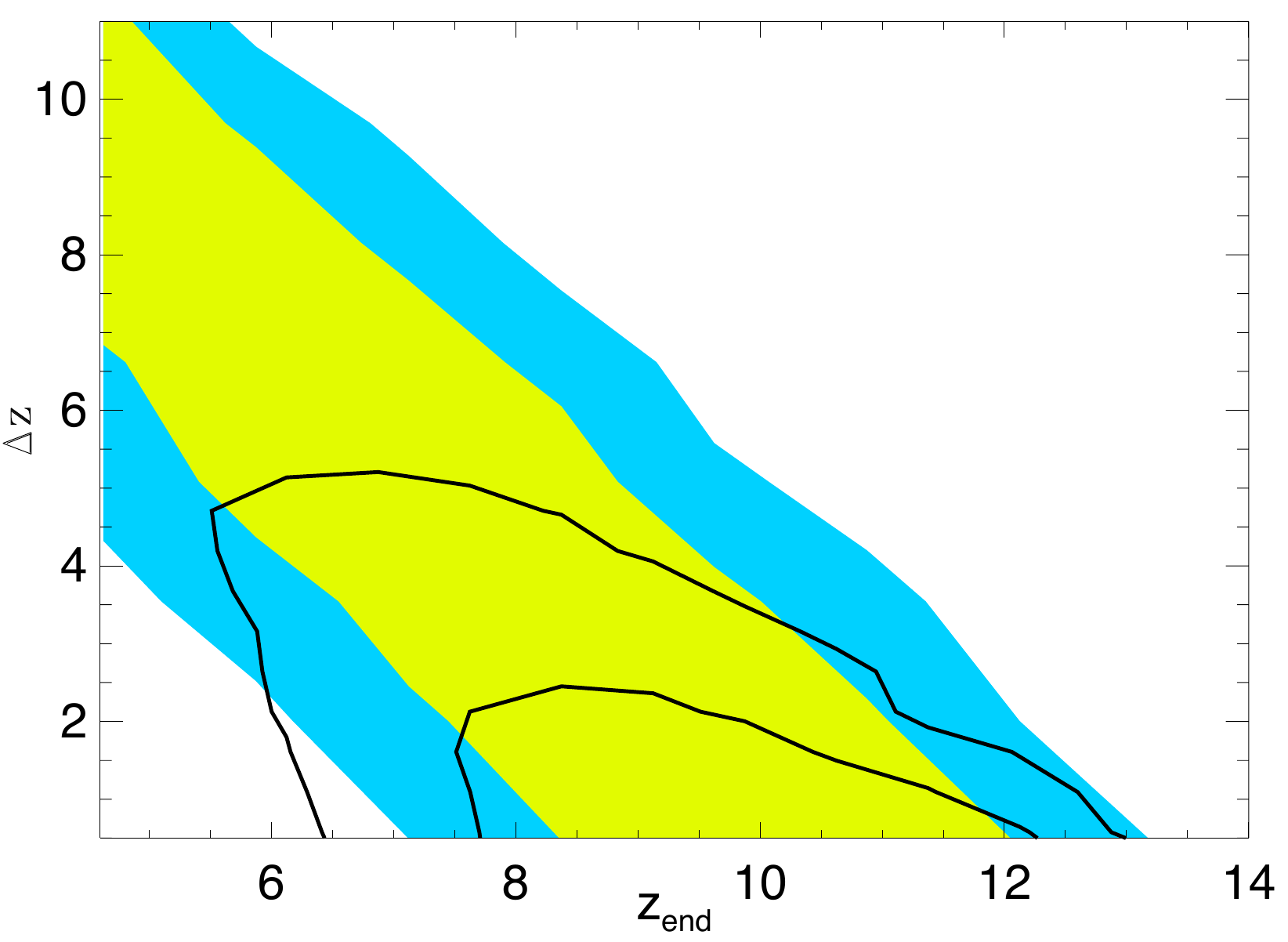}
\caption{Contour plot of $z_{\rm end}-\Delta$z plane showing 68\% and 95\% confidence regions. There is a degeneracy in the \wmap\ data between the end of reionization and the duration (\textbf{shaded contours}) which is broken by the addition of SPT data (\textbf{line contours}). 
The tSZ-CIB correlation is assumed to be zero. }
\label{fig:contour_zend-dz}
\end{figure}

As previously mentioned, the integrated optical depth constraint alone can not constrain the evolution of the ionized fraction as a function of time. Figure \ref{fig:contour_zend-dz} shows the 68/95\% confidence intervals in the $z_{\rm end}$-$\Delta z$ plane. There is a perfect degeneracy in the large-scale \wmap\ polarization data between the duration and end of reionization; a low redshift for the end of reionization can be made consistent with the data by increasing the duration and the redshift of the starting point of reionization. 
The SPT constraint on the patchy kSZ contribution breaks this degeneracy by constraining the duration, as shown for the case of no tSZ-CIB correlations in the inner contours of Figure \ref{fig:contour_zend-dz}. 

\begin{figure*}[t]
\centering
\includegraphics[width=18.2cm]{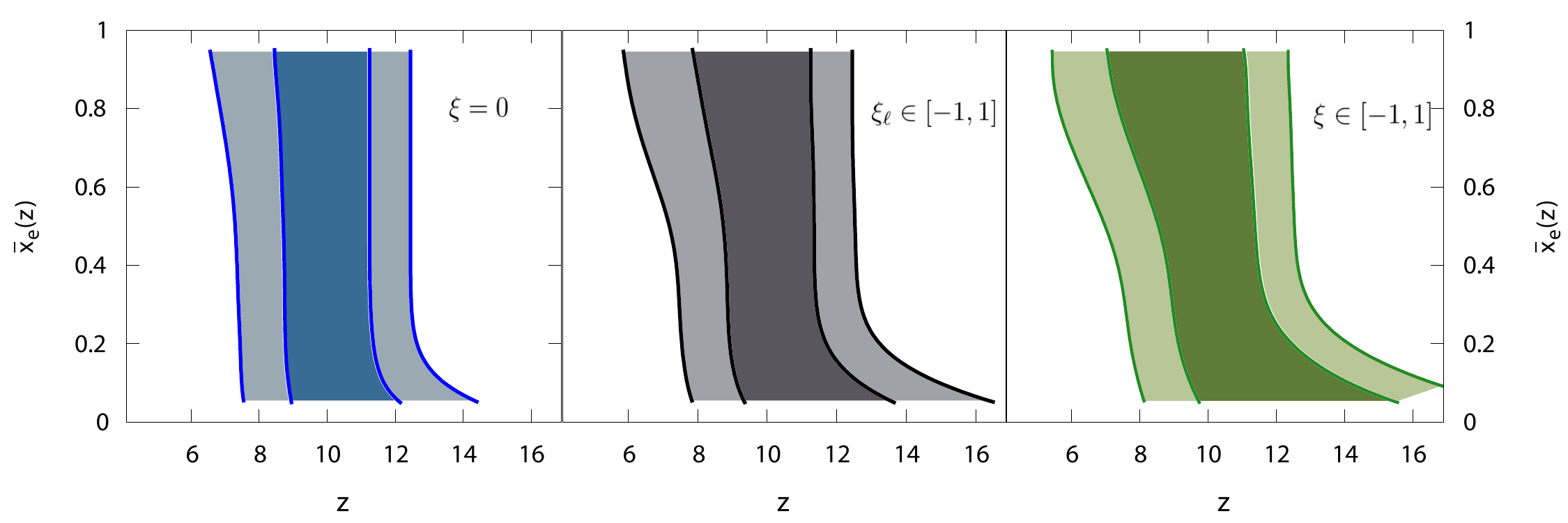}
\caption{CMB constraints on the evolution of the ionized fraction for three cases of the foreground modeling. The SPT+WMAP 68/95\% confidence intervals are indicated by the thick/thin curves. The \textbf{\textit{left panel}} shows the case without a tSZ-CIB correlation, the \textbf{\textit{center panel}} shows the free $\xi_\ell$ case, the \textbf{\textit{right panel}} shows the free flat $\xi$ case.}
\label{fig:xhiiz}
\end{figure*}

To obtain constraints on the evolution of the ionized fraction with redshift, we integrate the posterior distributions at each $\xhii$ to obtain the 68\% and 95\% likelihood intervals for that ionization fraction to fall within a certain redshift range. 
Again, we assume the CSF \hksz\ model for the post-reionization kSZ power. 
As indicated by Table  \ref{tab:pksz}, the effect of different homogeneous kSZ power is modest given our tight upper limits. 
The constraints on $\xhii$ from the addition of SPT to \wmap\ are shown for the three tSZ-CIB correlation variants in Figure \ref{fig:xhiiz}. 
The case with arbitrary, $\ell$-independent correlation (right panel, green lines) shows the weakest constraints. 
The case with rising correlation shape (center panel, black lines) gives intermediate constraints. 
The case without correlations (left panel, blue lines) shows the tightest constraint on the ionized fraction, with the error on the timing dominated by the \wmap\ optical depth uncertainty. 
The bulk of the epoch of reionization is confined to a narrow redshift interval from $z \simeq 12.5$ to $z \simeq 7$ at 95\% CL. 

\begin{figure*}[t]
\centering
\includegraphics[width=18.5cm]{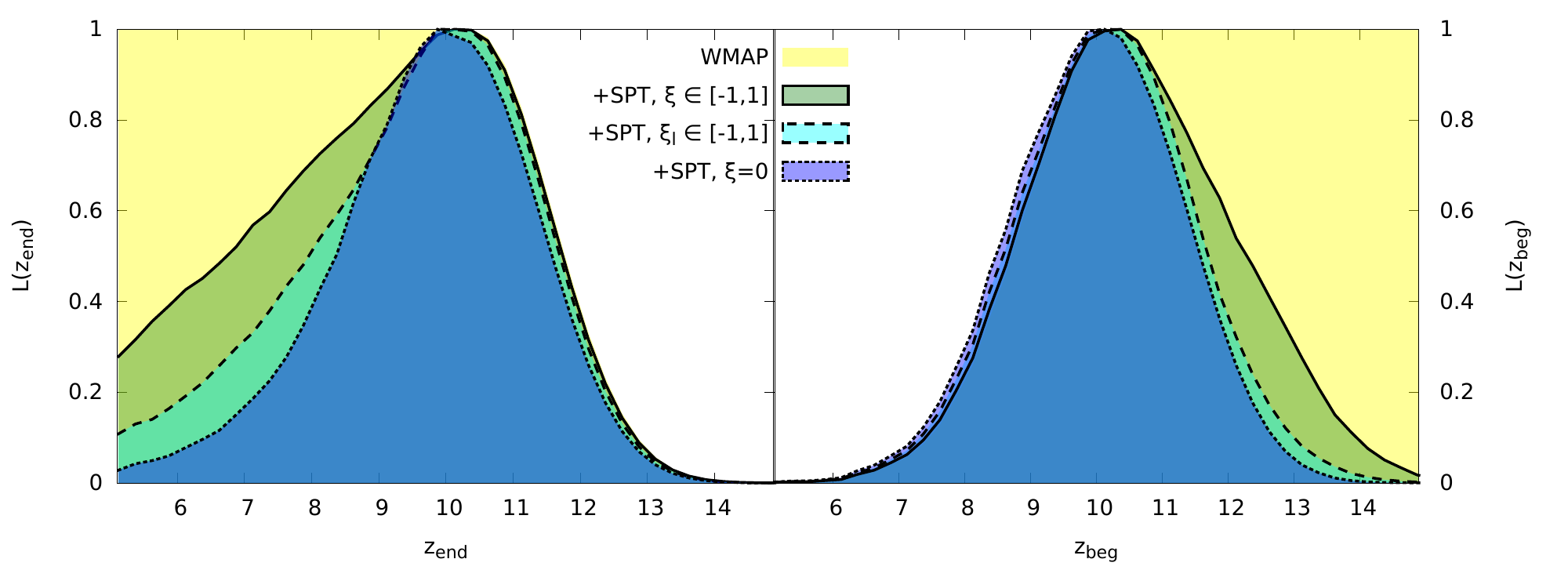}
\caption{
Likelihood functions for $z_{\rm end}$ ($x_e=0.99$, \textbf{\textit{left panel}}) and $z_{\rm beg}$ of reionization ($x_e=0.2$, \textbf{\textit{right panel}}). 
In both panels, \wmap\ data only requires that the end (beginning) of reionization will be after (before) the midpoint (yellow shading). 
The most conservative case with SPT data is shown by the green shaded region, solid line. 
This allows a tSZ-CIB correlation with an $\ell$-independent shape. 
With the cyan region we show results when allowing correlations with the $\ell$-dependent shape, dashed line. 
In the most restrictive case, we neglect correlations (blue shading, dotted line). 
We have assumed the CSF \hksz\ model for the post-reionization kSZ signal. }
\label{fig:zendzbeg_gammafree}
\end{figure*}

We next consider the end of reionization which has been the focus of most current reionization observations. 
Recall that we define the end of reionization as $\xhii=0.99$.
Constraints on $z_{\rm end}$ are shown in the  left panel of Figure \ref{fig:zendzbeg_gammafree} and tabulated in Table \ref{tab:summary}. 
This is equivalent to taking a slice through Figure \ref{fig:xhiiz} at $\xhii = 0.99$. 
The SPT data enable the first CMB constraint on $z_{\rm end}$. 
The SPT data disfavor a late end to reionization due to their preference for short durations. 
 For $\xi=0$, we find that having reionization end at $z_{\rm end} <  7.2$ is disfavored at 95\% CL (inner blue contour, dotted line). 
The constraint weakens considerably when an arbitrary, $\ell$-independent correlation is allowed (wider green contour, solid line).
However, in this most conservative interpretation of our data, reionization still concluded at $z_{\rm end}>5.8$ at 95\% CL. The rising-$\ell$ correlation shape yields an intermediate constraint of $z_{\rm end} > 6.4$ (cyan contour, dashed line). 
Note that while \wmap\ does not place a lower limit on the end of reionization, it trivially places an {\it upper} limit (the \wmap\ contour lies directly beneath the SPT contours on this side) by limiting the integrated optical depth. Note also that the innermost contour is shifted slightly to the left compared to the cases with tSZ-CIB correlation. This is because the tighter SPT upper limit on the combined kSZ means that short models ending at lower redshift are preferred because they entail a smaller homogeneous kSZ contribution. In other words, there is a slight tension between the SPT requirement for a small kSZ amplitude and the WMAP requirement for a relatively large integrated opacity.

Finally, we present constraints on when the first ionizing sources turn on and begin reionizing the universe. 
Again, we have defined the beginning of reionization by $\xhii=0.20$. 
The likelihood function for $z_{\rm beg}$ is shown in the right panel of Figure \ref{fig:zendzbeg_gammafree}. 
This is equivalent to taking a slice through Figure \ref{fig:xhiiz} at $\xhii = 0.20$. 
We find that the combination of \wmap\ and SPT data rule out an early onset of reionization at $z>12.1$ at 95\% confidence in the $\xi=0$ case. 
When allowing for a free, $\ell$-independent tSZ-CIB correlation, the 95\% confidence upper limit increases to $z_{\rm beg} \leq 13.1$. 
Again, the \wmap\ optical depth constraint leads only to a lower limit on the beginning of reionization, see above. Again the $\xi=0$ likelihood peaks at slightly lower redshift, for the same reason as in the $z_{\rm end}$ case. 

\begin{table}
\bc
\caption{Reionization results}
\begin{tabular}{cc|ccc}

\hline
\hline
 & &   $\xi=0$    &  $\xi_\ell\in[-1,1]$    &   $\xi\in[-1,1]  $   \\
 \hline
 \multirow{2}{*}{$\Delta z$} & (68\%)& 2.1 & 2.9 & 4.4 \\
&(95\%) &  {4.4} &{ 6.2}  &{7.9} \\
 \hline
  \multirow{2}{*}{$z_{\rm beg}$}  &(68\%)&10.7 & 10.9 & 11.2 \\
&(95\%) &  12.1 & 12.3 & 13.1 \\
\hline
  \multirow{2}{*}{$z_{\rm end}$}  &(68\%)&9.2 & 8.9 & 8.3 \\
&(95\%) &  7.2 & 6.4 & 5.8 \\
\end{tabular}
\label{tab:summary}
\ec
\tablecomments{Reionization results for different tSZ-CIB correlation assumptions. We have defined volume-weighted ionized fractions of 20\% and as the 99\% as the beginning $z_{\rm beg}$ and end $z_{\rm end}$ of reionization, respectively, and the duration $\Delta z$ as the redshift interval between the two.
We show 68\% and 95\% limits in each case as the likelihood surfaces are non-Gaussian. 
The first two rows ($\Delta z$ and  $z_{\rm beg}$) show upper limits, while the third row ($z_{\rm end})$ shows lower limits. 
The first column assumes the most restrictive model for the correlations, $\xi = 0$, and therefore leads to the tightest reionization limits.  
The second column allows free correlations with the rising model shape (see \S\ref{sec:cibtsz} and the solid line in Figure \ref{fig:tsz_cib_templates}) while the third column allows a free, $\ell$-independent tSZ-CIB correlation. 
The latter leads to the weakest limits on the epoch of reionization. 
}
\end{table}

\subsection{Forecasts for SPT full survey, Planck, and Herschel}

In the near future, we expect improved measurements of the kSZ power from the full SPT survey,  optical depth (from \planck), and CIB (from \herschel\, and \herschel/SPT cross-correlation analyses). 
The SPT survey of 2,500 square degrees (three times the area used in this work) was completed in November 2011. 
The \planck\ survey is ongoing and the first power spectrum results including the optical depth constraint from the large-scale E-mode polarization feature should be released in 2013. 
\herschel\ observations of the deepest 100 square degrees of the SPT survey will conclude in 2012, and should enable a detailed study of the tSZ-CIB correlation as well as the CIB in general. 

We estimate the improvement in the kSZ power constraint from the full SPT survey with 100 deg$^2$ of \herschel{{} overlap by running a Monte Carlo Markov chain  with simulated bandpowers and uncertainties. 
We assume a 1\% temperature calibration uncertainty and a 5\% beam FWHM uncertainty in the SPT frequency bands; we assume the \herschel{} data is used to create a CIB template map which will be subtracted from the SPT bands to reduce the CIB contribution. 
Given the wide frequency range spanned, we introduce one new parameter to allow for decorrelation in the CIB between frequency bands. 
Instead of a single $\beta$ in equation \ref{eqn:cib}, we assume a distribution of $\beta$ in the galaxies of N($\beta$,$\sigma_\beta$). 
We set a uniform prior of $\sigma_\beta^2 \in [0,0.35]$; the upper edge is chosen such that the correlation between 150 and 220\,GHz is at least 95\%. 
For the full SPT survey without \herschel{}, we find the kSZ constraint improves proportionally to the reduced SPT  bandpower uncertainties; the uncertainties on the bandpowers and kSZ power are reduced by a factor of $\sqrt{3}$.  The combination of SPT and \herschel{{} leads to a factor of 6 improvement in the kSZ constraints. 
Assuming the post-reionization kinetic SZ contribution is known, we find that the combination of SPT and \herschel\ data should be able to positively identify extended reionization at 95\% confidence if $\Delta z\geq 2$. Using the \planck\ published sensitivity numbers leads to a forecasted constraint on the optical depth of $\Delta \tau \simeq 0.005$ \citep{Zaldarriaga:2008ap}.
This is almost three times better than the current WMAP7+SPT $\tau = 0.085\pm 0.014$ constraint \citep{keisler11}.
Figure \ref{fig:forecast} shows constraint forecasts for the end of reionization. 
Our current constraints with an uncertain, $\ell$-independent, tSZ-CIB correlation is reproduced for comparison in the green contour, solid line. 
We show three variations on the predicted $z_{\rm end}$ likelihood function. 
The cyan contour (dotted line) shows constraints centered on the WMAP7+SPT optical depth if the combined future data set continues to set only a stricter upper limit on the kSZ power. 
The predicted $z_{\rm end}$ tightens significantly around $z_{\rm end} \simeq 10.25$ and models that end at $z < 8$ would be ruled out at 95\% confidence. 
At the same $\tau$ value, a measured patchy kSZ amplitude of $2\, \mu K^2$, corresponding to a reionization duration of $\Delta z\simeq 4$ and in good agreement with the flat correlation at the preferred SPT value of $\xi = -0.18$ leads to the blue contour (dashed line). 
Reducing the central optical depth to the $\sim$1~$\sigma$ lower bound of the \wmap\ measurement with zero tSZ-CIB correlation leads to the red contour (dot-dashed line). 
The latter two cases have similar effects on the $z_{\rm end}$ constraints. 
The predicted $z_{\rm end}$ tightensÊaround $z_{\rm end} \simeq 8-8.5$ and models that end at $z \lesssim 6.5$ would be disfavored at $\geq$ 95\% confidence.

\begin{figure}[t]
\centering
\includegraphics[width=9cm]{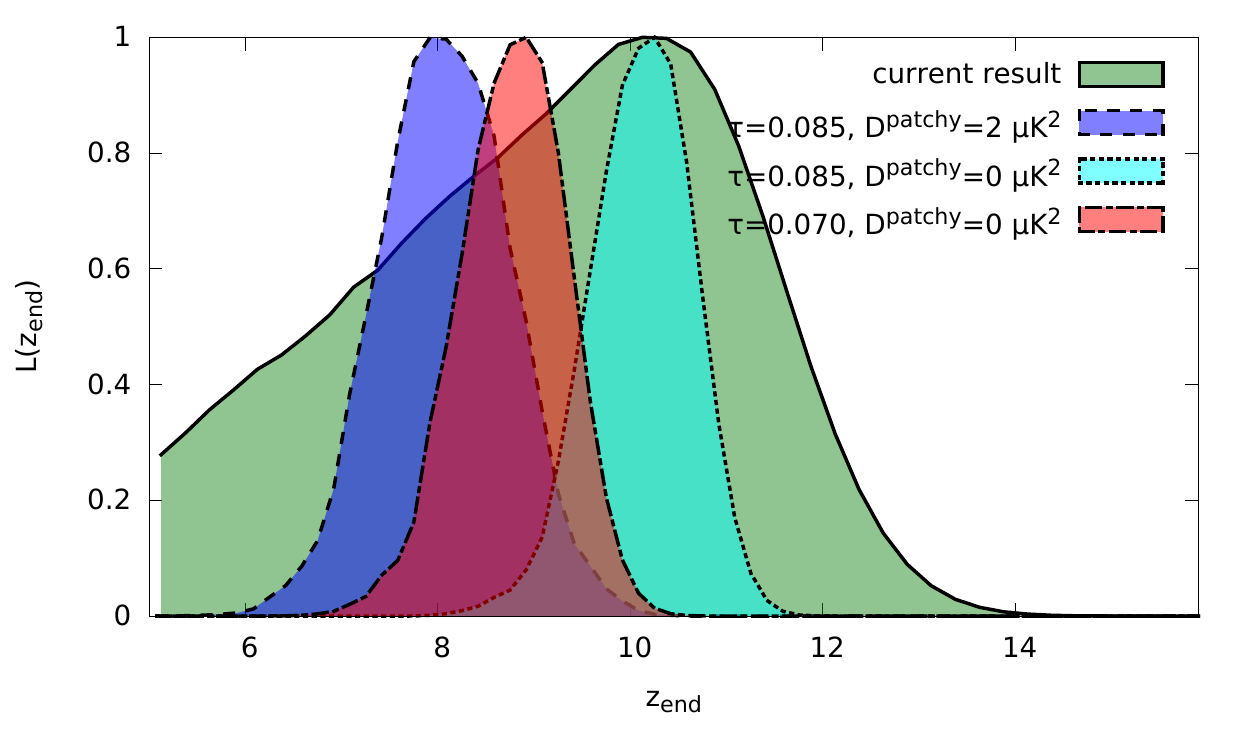}
\caption{Forecasted $z_{\rm end}$ likelihood functions for future SPT+\planck+\herschel\ data.
The \textbf{blue region, dashed line} shows the predicted  likelihood function assuming zero tSZ-CIB correlation and the median WMAP7 optical depth. 
We explore the effects of shifting the best fit values for $\tau$ to the WMAP7 lower 1~$\sigma$ bounds in the \textbf{red region, dot-dashed line}, or alternatively of shifting the $\ell$-independent tSZ-CIB correlation to the R11 preference of $\xi=-0.18$ in the  \textbf{cyan region, dotted line}. 
The current conservative foreground modeling constraint is shown by the  \textbf{green region, solid line}.}
\label{fig:forecast}
\end{figure}

\section{Discussion and Conclusion}
\label{sec:discussion}

In this work, we produce the first constraints on the evolution of the ionized fraction during the epoch of reionization using small scale CMB observations. 
To this end, we have presented a framework based on efficient reionization simulations to calculate joint constraints on $\xhii$ from measurements of the total optical depth and kSZ power. 
We have applied this method to new observations of the kSZ power from the SPT (R11) and published optical depth results from WMAP7 \citep{komatsu11} to probe the beginning, end, and duration of the epoch of reionization. 

We find that the SPT kSZ measurement implies a short reionization duration. 
We show that the SPT kSZ constraint is sensitive to modeling assumptions about the poorly known tSZ-CIB correlation. 
Assuming this correlation to be negligible leads to the strongest limits on the epoch of reionization duration. 
Conversely, assuming no outside knowledge of the correlation leads to the most conservative results. 
The 95\% upper limits on the duration are $\Delta z<4.4$ under the assumption of no correlations and $\Delta z < 7.9$ for the most conservative assumptions. 
As a result, the SPT data combined with the WMAP optical depth constraint rule out reionization models that end very late or begin very early. 
With $\ell$-independent correlations, the epoch of reionization ends at $z>5.8$ and begins at $z < 13.1$ at 95\% CL.

\begin{figure*}[t]
\centering
\vspace{-1cm}
\hspace*{-2cm}
\includegraphics[width=16cm,angle=-90]{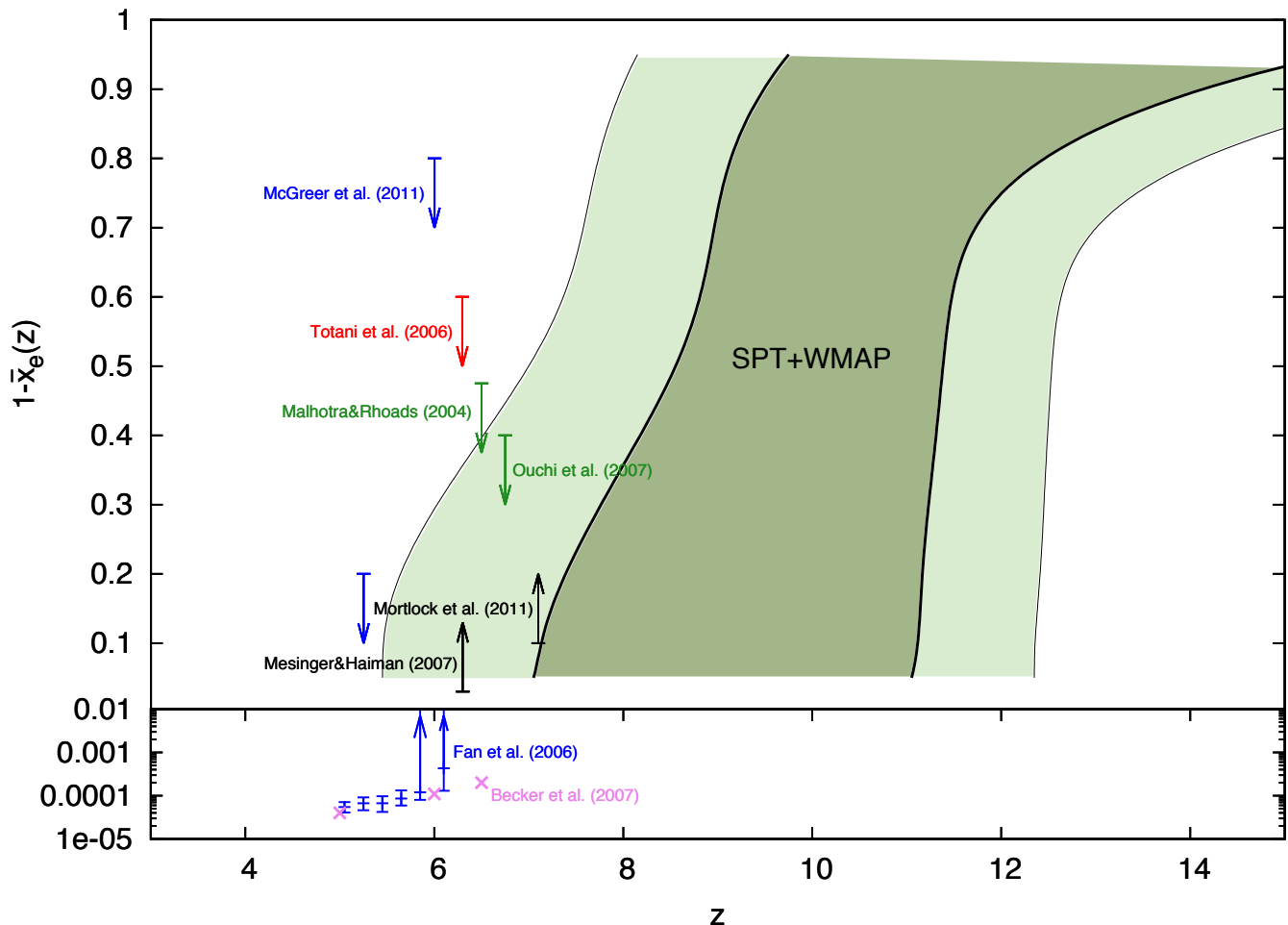}
\vspace{-1.8cm}
\caption{CMB constraint on the redshift evolution of the mean neutral fraction for the (most conservative) free amplitude, $\ell$-independent tSZ-CIB correlation case. The SPT+WMAP 68/95\% confidence ranges are indicated by the thick/thin curves and the dark/light green shading.
We also show other constraints on the neutral fraction based on quasar spectra (blue and violet constraints as well as black lower limits), a gamma ray burst (red upper limit), and Ly$\alpha$ emitters (green upper limits).}
\label{fig:summary}
\end{figure*}

The end of reionization has been the subject of intense theoretical interest and observational effort. 
In Figure \ref{fig:summary}, we contrast the most conservative CMB-derived constraint on the {\it neutral} fraction $\xhi(z)$  (where $\xhii \leq 1$) with other data. 
We choose to show $\xhi$ instead of $\xhii$ as this is more natural for the other datasets. Specifically, we are allowing an $\ell$-independent tSZ-CIB correlation and are assuming the best-guess CSF \hksz\ model for the post-reionization kSZ signal. 
The dark and light green contours are the 1 and 2~$\sigma$ likelihood contours respectively from the CMB data. 
Also shown in the upper panel are external constraints on reionization from the Ly-$\alpha$ forest \citep{McGreer2011}, Ly-$\alpha$ emitters (LAEs) \citep{MalhotraRhoads2004,Ouchi2010}, a gamma ray burst GRB 050904 \citep{Totani2006}, and quasar proximity regions \citep{MesingerHaiman2007,Mortlock:2011va}. In the lower panel, we plot constraints on the residual neutral fraction after reionization from the Ly-$\alpha$ forest \citep{Fan2006,BeckerRS2007}.
The previously published data constrain the relatively narrow interval of redshifts $\simeq 5-7$. 
Most results are upper limits, meaning they are consistent with reionization having concluded much earlier. This is also true for the Ly-$\alpha$ forest points in the lower panel which correspond to very low neutral fractions \citep{Fan2006,BeckerRS2007}.
The SPT+WMAP7 data narrows in on a  previously unexplored region of the $(z,1-\xhii)$-plane.  

Two constraints are in mild tension with our results. These are the {\em lower limits} on $\xhi$ obtained from possible detections of damping wing absorption in quasar proximity zones by \citet{MesingerHaiman2004,MesingerHaiman2007}, and to a lesser extent \citet{Mortlock:2011va,Bolton:2011vb}. 
Additional proximity zone spectra, further modeling (e.g., \citealt{BoltonHaehnelt2007a,Maselli2007,LidzOF2006}), and constraints from other datasets should clarify the ionization state of the IGM at $z=6-7$.

In the future, it would be interesting to investigate whether there is tension with observations of the Ly-$\alpha$ forest {\em after reionization}. 
In particular, measurements of the mean transmitted flux through
the Ly-$\alpha$ forest at $z \sim 5$ indicate that the ionizing sources
emit only a few ionizing photons per hydrogen atom per Hubble time \citep{MiraldaEscude:2002yd,BoltonHaehnelt2007b}. 
Since after accounting for recombinations, a couple of photons per atom are required to complete reionization, these measurements suggest that reionization is a  prolonged process.

The results presented here have implications for the proposed study of redshifted 21 cm emission from the hyperfine transition of neutral hydrogen.
Observations of the hyperfine transition can potentially provide detailed three-dimensional information about the evolution and morphology of the reionization process \citep[e.g.,][]{ZaldarriagaFH2004}.  
In the next few years, an ambitious new generation of telescopes will begin collecting data to detect this signal.  
These efforts include the MWA \citep{BowmanMH2005}, LOFAR \citep{Harker2011}, GMRT \citep{Pen2009}, PAPER \citep{Parsons2009} and SKA \citep{Johnston:2008hp}.
Improved constraints on the history of reionization can help
these experiments select the optimal frequency range in which to focus
their observations.

We also note that our results are consistent with the recent result based on the redshifted hydrogen hyperfine transition ``global step'' experiment EDGES, which produced a lower limit of $\Delta z\geq 0.06$ \citep{bowman10}.
Our results are good news for upcoming experiments that hope to detect a redshifted 21 cm absorption feature sourced by substantial UV coupling prior to X-ray heating during the onset of reionization \citep{Pritchard:2010pa,Bowman:2007su,Bowman:2009yc,Burns2011,Harker2011}. 
The combination of SPT and WMAP7 data limits the early stages of reionization to $z_{\rm beg} \le$ 13.1 at the 95\% confidence level, meaning that an ionized fraction of 20\% was reached later. While the hyperfine structure absorption feature is expected at somewhat lower ionization fractions, our result indicates that the redshifted 21cm line radiation from this epoch will appear at frequencies higher than those in models with higher $z_{\rm beg}$.  There is less galactic synchrotron radiation and radio interference at these higher frequencies.

With a wealth of new data  on the horizon, this is an exciting period for CMB reionization constraints. 
The SPT survey of 6\% of the sky is now complete. 
The kSZ constraints from the full survey should be approximately $\sqrt{3}$ times better than those reported here; kSZ constraints will improve further with the deep SPTpol and ACTpol maps. 
The \herschel\ satellite is making unprecedented maps of the CIB, including a 100\,deg$^2$ overlap with the deepest 100 square degrees of the SPT survey. Further sub-mm surveys have started observations or will do so in the near future (e.g., \citealt{alma,ccat,scuba2}).
The cross-correlation analysis of \herschel\ and SPT will provide
information on the dominant foreground uncertainty for the kSZ power:
tSZ-CIB correlations.
Finally, the Planck survey is ongoing; the first power spectrum results are expected in 2013. 
The Planck data should improve the optical depth constraint by a factor of 3. 
The combination of these data sets should be able to positively identify extended reionization at 95\% confidence if $\Delta z\geq 2$, further illuminating the epoch of reionization.  

\acknowledgments

We thank Matt George, Guilaine Lagache, Cien Shang, David Spergel, Marco Viero, and Matias Zaldarriaga for useful discussions.
The South Pole Telescope is supported by the National Science
Foundation through grants ANT-0638937 and ANT-0130612.  Partial
support is also provided by the NSF Physics Frontier Center grant
PHY-0114422 to the Kavli Institute of Cosmological Physics at the
University of Chicago, the Kavli Foundation and the Gordon and Betty
Moore Foundation.
O. Zahn acknowledges support from a Berkeley Center for Cosmological Physics fellowship. The McGill group acknowledges funding from the National
Sciences and Engineering Research Council of Canada,
Canada Research Chairs program, and
the Canadian Institute for Advanced Research. 
R. Keisler acknowledges support from NASA Hubble Fellowship grant HF-51275.01.
B.A. Benson is supported by a KICP Fellowship.
M. Dobbs acknowledges support from an Alfred P. Sloan Research Fellowship.
L. Shaw acknowledges the support of Yale University and NSF grant AST-1009811.
M. Millea and L. Knox acknowledge the support of NSF grant 0709498.
This research used resources of the National Energy Research Scientific Computing Center, 
 which is supported by the Office of Science of the U.S. Department of Energy under Contract No. DE-AC02-05CH11231. Some of the results in this paper have been derived using the 
HEALPix package \citep{gorski05}. We acknowledge the use of the Legacy Archive for Microwave Background Data Analysis (LAMBDA). Support for LAMBDA is provided by the NASA Office 
of Space Science. We also acknowledge usage of the FFTW and TeXShop software packages. 

\bibliographystyle{apj}
\bibliography{spt}

\end{document}

%% file: definitions.tex
\newcommand\lsim{\mathrel{\rlap{\lower4pt\hbox{\hskip1pt$\sim$}}
        \raise1pt\hbox{$<$}}}
\newcommand\gsim{\mathrel{\rlap{\lower4pt\hbox{\hskip1pt$\sim$}}
        \raise1pt\hbox{$>$}}}

\newcommand{\erfc}{{\rm erfc}}

\def\n{{\bf  \hat n}}

\def\hmsun{\; h^{-1} \; M_{\odot}}

\def\xhii{{\bar x}_e}
\def\xhi{1-{\bar x}_e}

\newcommand{\bma}{\begin{math}}
\newcommand{\ema}{\end{math}}
\newcommand{\beq}{\begin{equation}}
\newcommand{\eeq}{\end{equation}}
\newcommand{\beqa}{\begin{eqnarray}}
\newcommand{\eeqa}{\end{eqnarray}}
\newcommand{\bc}{\begin{center}}
\newcommand{\ec}{\end{center}} 
\newcommand{\bit}{\begin{itemize}}
\newcommand{\eit}{\end{itemize}}

\newcommand{\herschel}{\textit{Herschel}}
\newcommand{\planck}{\textit{Planck}}
\newcommand{\wmap}{\textit{WMAP}}
\newcommand{\planckhfi}{\textit{Planck}/HFI}
\newcommand{\hksz}{{\rm homogeneous kSZ}}
\newcommand{\Dhksz}{\ensuremath{{\rm D^{homog}_{3000}}}}

%% file: ms.bbl
\begin{thebibliography}{116}
\expandafter\ifx\csname natexlab\endcsname\relax\def\natexlab#1{#1}\fi

\bibitem[{{Addison} {et~al.}(2012){Addison}, {Dunkley}, {Hajian}, {Viero},
  {Bond}, {Das}, {Devlin}, {Halpern}, {Hincks}, {Hlozek}, {Marriage},
  {Moodley}, {Page}, {Reese}, {Scott}, {Spergel}, {Staggs}, \&
  {Wollack}}]{addison11}
{Addison}, G.~E., {et~al.} 2012, \apj, 752, 120

\bibitem[{{Altay} {et~al.}(2008){Altay}, {Croft}, \& {Pelupessy}}]{AltayCP2008}
{Altay}, G., {Croft}, R.~A.~C., \& {Pelupessy}, I. 2008, \mnras, 386, 1931

\bibitem[{{Alvarez} {et~al.}(2009){Alvarez}, {Busha}, {Abel}, \&
  {Wechsler}}]{Alvarez2009}
{Alvarez}, M.~A., {Busha}, M., {Abel}, T., \& {Wechsler}, R.~H. 2009, \apjl,
  703, L167

\bibitem[{{Aubert} \& {Teyssier}(2008)}]{AubertTeyssier2008}
{Aubert}, D., \& {Teyssier}, R. 2008, \mnras, 387, 295

\bibitem[{Barkana \& Loeb(2001)}]{Barkana:2000fd}
Barkana, R., \& Loeb, A. 2001, Phys.Rept., 349, 125

\bibitem[{{Battaglia} {et~al.}(2011){Battaglia}, {Bond}, {Pfrommer}, \&
  {Sievers}}]{battaglia11}
{Battaglia}, N., {Bond}, J.~R., {Pfrommer}, C., \& {Sievers}, J.~L. 2011, ArXiv
  e-prints

\bibitem[{{Becker} {et~al.}(2007){Becker}, {Rauch}, \&
  {Sargent}}]{BeckerRS2007}
{Becker}, G.~D., {Rauch}, M., \& {Sargent}, W.~L.~W. 2007, \apj, 662, 72

\bibitem[{{Bolton} \& {Haehnelt}(2007{\natexlab{a}})}]{BoltonHaehnelt2007a}
{Bolton}, J.~S., \& {Haehnelt}, M.~G. 2007{\natexlab{a}}, \mnras, 381, L35

\bibitem[{{Bolton} \& {Haehnelt}(2007{\natexlab{b}})}]{BoltonHaehnelt2007b}
---. 2007{\natexlab{b}}, \mnras, 382, 325

\bibitem[{{Bolton} {et~al.}(2011){Bolton}, {Haehnelt}, {Warren}, {Hewett},
  {Mortlock}, {Venemans}, {McMahon}, \& {Simpson}}]{Bolton:2011vb}
{Bolton}, J.~S., {Haehnelt}, M.~G., {Warren}, S.~J., {Hewett}, P.~C.,
  {Mortlock}, D.~J., {Venemans}, B.~P., {McMahon}, R.~G., \& {Simpson}, C.
  2011, \mnras, 416, L70

\bibitem[{Bond {et~al.}(1991)Bond, Cole, Efstathiou, \& Kaiser}]{Bond:1990iw}
Bond, J., Cole, S., Efstathiou, G., \& Kaiser, N. 1991, Astrophys.J., 379, 440

\bibitem[{{Bowman} {et~al.}(2005){Bowman}, {Morales}, \&
  {Hewitt}}]{BowmanMH2005}
{Bowman}, J.~D., {Morales}, M.~F., \& {Hewitt}, J.~N. 2005, in Bulletin of the
  American Astronomical Society, Vol.~37, Bulletin of the American Astronomical
  Society, 1217--+

\bibitem[{Bowman \& Rogers(2010)}]{bowman10}
Bowman, J.~D., \& Rogers, A. E.~E. 2010, Nature, 468, 796

\bibitem[{{Bowman} {et~al.}(2008{\natexlab{a}}){Bowman}, {Rogers}, \&
  {Hewitt}}]{Bowman:2009yc}
{Bowman}, J.~D., {Rogers}, A.~E.~E., \& {Hewitt}, J.~N. 2008{\natexlab{a}}, in
  American Institute of Physics Conference Series, Vol. 1035, The Evolution of
  Galaxies Through the Neutral Hydrogen Window, ed. R.~{Minchin} \&
  E.~{Momjian}, 87--89

\bibitem[{{Bowman} {et~al.}(2008{\natexlab{b}}){Bowman}, {Rogers}, \&
  {Hewitt}}]{Bowman:2007su}
{Bowman}, J.~D., {Rogers}, A.~E.~E., \& {Hewitt}, J.~N. 2008{\natexlab{b}},
  \apj, 676, 1

\bibitem[{Bryan \& Norman(1998)}]{Bryan:1997dn}
Bryan, G., \& Norman, M. 1998, Astrophys.J., 495, 80

\bibitem[{{Burns} {et~al.}(2012){Burns}, {Lazio}, {Bale}, {Bowman}, {Bradley},
  {Carilli}, {Furlanetto}, {Harker}, {Loeb}, \& {Pritchard}}]{Burns2011}
{Burns}, J.~O., {et~al.} 2012, Advances in Space Research, 49, 433

\bibitem[{{Cen}(2003)}]{Cen2003}
{Cen}, R. 2003, \apj, 591, 12

\bibitem[{{Choudhury} {et~al.}(2009){Choudhury}, {Haehnelt}, \&
  {Regan}}]{ChoudhuryHR2009}
{Choudhury}, T.~R., {Haehnelt}, M.~G., \& {Regan}, J. 2009, \mnras, 394, 960

\bibitem[{{Ciardi} {et~al.}(2003){Ciardi}, {Ferrara}, \& {White}}]{Ciardi2003}
{Ciardi}, B., {Ferrara}, A., \& {White}, S.~D.~M. 2003, \mnras, 344, L7

\bibitem[{{Crociani} {et~al.}(2011){Crociani}, {Mesinger}, {Moscardini}, \&
  {Furlanetto}}]{Crociani:2010qe}
{Crociani}, D., {Mesinger}, A., {Moscardini}, L., \& {Furlanetto}, S. 2011,
  \mnras, 411, 289

\bibitem[{{Dwek} \& {Arendt}(1998)}]{dwek98}
{Dwek}, E., \& {Arendt}, R.~G. 1998, \apjl, 508, L9

\bibitem[{{Fan} {et~al.}(2006{\natexlab{a}}){Fan}, {Carilli}, \&
  {Keating}}]{fan06}
{Fan}, X., {Carilli}, C.~L., \& {Keating}, B. 2006{\natexlab{a}}, \araa, 44,
  415

\bibitem[{{Fan} {et~al.}(2006{\natexlab{b}}){Fan}, {Strauss}, {Becker},
  {White}, {Gunn}, {Knapp}, {Richards}, {Schneider}, {Brinkmann}, \&
  {Fukugita}}]{Fan2006}
{Fan}, X., {et~al.} 2006{\natexlab{b}}, \aj, 132, 117

\bibitem[{{Finlator} {et~al.}(2009){Finlator}, {{\"O}zel}, \&
  {Dav{\'e}}}]{FinlatorOD2009}
{Finlator}, K., {{\"O}zel}, F., \& {Dav{\'e}}, R. 2009, \mnras, 393, 1090

\bibitem[{{Fixsen} {et~al.}(1998){Fixsen}, {Dwek}, {Mather}, {Bennett}, \&
  {Shafer}}]{fixsen98}
{Fixsen}, D.~J., {Dwek}, E., {Mather}, J.~C., {Bennett}, C.~L., \& {Shafer},
  R.~A. 1998, \apj, 508, 123

\bibitem[{{Furlanetto} \& {Oh}(2005)}]{FurlanettoOh2005}
{Furlanetto}, S.~R., \& {Oh}, S.~P. 2005, \mnras, 363, 1031

\bibitem[{{Furlanetto} {et~al.}(2004){Furlanetto}, {Zaldarriaga}, \&
  {Hernquist}}]{Furlanetto2004}
{Furlanetto}, S.~R., {Zaldarriaga}, M., \& {Hernquist}, L. 2004, \apj, 613, 1

\bibitem[{{Furlanetto} {et~al.}(2006){Furlanetto}, {Zaldarriaga}, \&
  {Hernquist}}]{FurlanettoZH2006}
---. 2006, \mnras, 365, 1012

\bibitem[{{Geil} \& {Wyithe}(2008)}]{GeilWyithe2008}
{Geil}, P.~M., \& {Wyithe}, J.~S.~B. 2008, \mnras, 386, 1683

\bibitem[{{George} {et~al.}(2011){George}, {Leauthaud}, {Bundy}, {Finoguenov},
  {Tinker}, {Lin}, {Mei}, {Kneib}, {Aussel}, {Behroozi}, {Busha}, {Capak},
  {Coccato}, {Covone}, {Faure}, {Fiorenza}, {Ilbert}, {Le Floc'h}, {Koekemoer},
  {Tanaka}, {Wechsler}, \& {Wolk}}]{george11}
{George}, M.~R., {et~al.} 2011, \apj, 742, 125

\bibitem[{{Gnedin}(2000)}]{Gnedin2000}
{Gnedin}, N.~Y. 2000, \apj, 542, 535

\bibitem[{{G{\'o}rski} {et~al.}(2005){G{\'o}rski}, {Hivon}, {Banday},
  {Wandelt}, {Hansen}, {Reinecke}, \& {Bartelmann}}]{gorski05}
{G{\'o}rski}, K.~M., {Hivon}, E., {Banday}, A.~J., {Wandelt}, B.~D., {Hansen},
  F.~K., {Reinecke}, M., \& {Bartelmann}, M. 2005, \apj, 622, 759

\bibitem[{{Gruzinov} \& {Hu}(1998)}]{Gruzinov:1998u}
{Gruzinov}, A., \& {Hu}, W. 1998, \apj, 508, 435

\bibitem[{{Haiman} \& {Cen}(2005)}]{HaimanCen2005}
{Haiman}, Z., \& {Cen}, R. 2005, \apj, 623, 627

\bibitem[{{Hall} {et~al.}(2010){Hall}, {Knox}, {Reichardt}, {Ade}, {Aird},
  {Benson}, {Bleem}, {Carlstrom}, {Chang}, {Cho}, {Crawford}, {Crites}, {de
  Haan}, {Dobbs}, {George}, {Halverson}, {Holder}, {Holzapfel}, {Hrubes},
  {Joy}, {Keisler}, {Lee}, {Leitch}, {Lueker}, {McMahon}, {Mehl}, {Meyer},
  {Mohr}, {Montroy}, {Padin}, {Plagge}, {Pryke}, {Ruhl}, {Schaffer}, {Shaw},
  {Shirokoff}, {Spieler}, {Staniszewski}, {Stark}, {Switzer}, {Vanderlinde},
  {Vieira}, {Williamson}, \& {Zahn}}]{hall10}
{Hall}, N.~R., {et~al.} 2010, \apj, 718, 632

\bibitem[{{Harker} {et~al.}(2012){Harker}, {Pritchard}, {Burns}, \&
  {Bowman}}]{Harker2011}
{Harker}, G.~J.~A., {Pritchard}, J.~R., {Burns}, J.~O., \& {Bowman}, J.~D.
  2012, \mnras, 419, 1070

\bibitem[{{Holland} {et~al.}(2006){Holland}, {MacIntosh}, {Fairley}, {Kelly},
  {Montgomery}, {Gostick}, {Atad-Ettedgui}, {Ellis}, {Robson}, {Hollister},
  {Woodcraft}, {Ade}, {Walker}, {Irwin}, {Hilton}, {Duncan}, {Reintsema},
  {Walton}, {Parkes}, {Dunare}, {Fich}, {Kycia}, {Halpern}, {Scott}, {Gibb},
  {Molnar}, {Chapin}, {Bintley}, {Craig}, {Chylek}, {Jenness}, {Economou}, \&
  {Davis}}]{scuba2}
{Holland}, W., {et~al.} 2006, in Society of Photo-Optical Instrumentation
  Engineers (SPIE) Conference Series, Vol. 6275, Society of Photo-Optical
  Instrumentation Engineers (SPIE) Conference Series

\bibitem[{Hultman~Kramer {et~al.}(2006)Hultman~Kramer, Haiman, \&
  Oh}]{HultmanKramer:2006dn}
Hultman~Kramer, R., Haiman, Z., \& Oh, S.~P. 2006, Astrophys. J., 649, 570

\bibitem[{{Iliev} {et~al.}(2006){Iliev}, {Mellema}, {Pen}, {Merz}, {Shapiro},
  \& {Alvarez}}]{Iliev2006b}
{Iliev}, I.~T., {Mellema}, G., {Pen}, U.-L., {Merz}, H., {Shapiro}, P.~R., \&
  {Alvarez}, M.~A. 2006, \mnras, 369, 1625

\bibitem[{Iliev {et~al.}(2007)Iliev, Mellema, Shapiro, \& Pen}]{Iliev:2006sw}
Iliev, I.~T., Mellema, G., Shapiro, P.~R., \& Pen, U.-L. 2007, Mon. Not. Roy.
  Astron. Soc., 376, 534

\bibitem[{{Iliev} {et~al.}(2007){Iliev}, {Pen}, {Bond}, {Mellema}, \&
  {Shapiro}}]{iliev06}
{Iliev}, I.~T., {Pen}, U.-L., {Bond}, J.~R., {Mellema}, G., \& {Shapiro}, P.~R.
  2007, \apj, 660, 933

\bibitem[{{Johnston} {et~al.}(2008){Johnston}, {Taylor}, {Bailes}, {Bartel},
  {Baugh}, {Bietenholz}, {Blake}, {Braun}, {Brown}, {Chatterjee}, {Darling},
  {Deller}, {Dodson}, {Edwards}, {Ekers}, {Ellingsen}, {Feain}, {Gaensler},
  {Haverkorn}, {Hobbs}, {Hopkins}, {Jackson}, {James}, {Joncas}, {Kaspi},
  {Kilborn}, {Koribalski}, {Kothes}, {Landecker}, {Lenc}, {Lovell}, {Macquart},
  {Manchester}, {Matthews}, {McClure-Griffiths}, {Norris}, {Pen}, {Phillips},
  {Power}, {Protheroe}, {Sadler}, {Schmidt}, {Stairs}, {Staveley-Smith},
  {Stil}, {Tingay}, {Tzioumis}, {Walker}, {Wall}, \&
  {Wolleben}}]{Johnston:2008hp}
{Johnston}, S., {et~al.} 2008, Experimental Astronomy, 22, 151

\bibitem[{{Kashikawa} {et~al.}(2006){Kashikawa}, {Shimasaku}, {Malkan}, {Doi},
  {Matsuda}, {Ouchi}, {Taniguchi}, {Ly}, {Nagao}, {Iye}, {Motohara},
  {Murayama}, {Murozono}, {Nariai}, {Ohta}, {Okamura}, {Sasaki}, {Shioya}, \&
  {Umemura}}]{Kashikawa2006}
{Kashikawa}, N., {et~al.} 2006, \apj, 648, 7

\bibitem[{{Keisler} {et~al.}(2011){Keisler}, {Reichardt}, {Aird}, {Benson},
  {Bleem}, {Carlstrom}, {Chang}, {Cho}, {Crawford}, {Crites}, {de Haan},
  {Dobbs}, {Dudley}, {George}, {Halverson}, {Holder}, {Holzapfel}, {Hoover},
  {Hou}, {Hrubes}, {Joy}, {Knox}, {Lee}, {Leitch}, {Lueker}, {Luong-Van},
  {McMahon}, {Mehl}, {Meyer}, {Millea}, {Mohr}, {Montroy}, {Natoli}, {Padin},
  {Plagge}, {Pryke}, {Ruhl}, {Schaffer}, {Shaw}, {Shirokoff}, {Spieler},
  {Staniszewski}, {Stark}, {Story}, {van Engelen}, {Vanderlinde}, {Vieira},
  {Williamson}, \& {Zahn}}]{keisler11}
{Keisler}, R., {et~al.} 2011, \apj, submitted, arXiv:1105.3182

\bibitem[{{Knox} {et~al.}(1998){Knox}, {Scoccimarro}, \& {Dodelson}}]{knox98}
{Knox}, L., {Scoccimarro}, R., \& {Dodelson}, S. 1998, Physical Review Letters,
  81, 2004

\bibitem[{{Kogut} {et~al.}(2003){Kogut}, {Spergel}, {Barnes}, {Bennett},
  {Halpern}, {Hinshaw}, {Jarosik}, \& {Limon}}]{Kogut2003}
{Kogut}, A., {Spergel}, D.~N., {Barnes}, C., {Bennett}, C.~L., {Halpern}, M.,
  {Hinshaw}, G., {Jarosik}, N., \& {Limon}, M. 2003, \apjs, 148, 161

\bibitem[{{Komatsu} {et~al.}(2011){Komatsu}, {Smith}, {Dunkley}, {Bennett},
  {Gold}, {Hinshaw}, {Jarosik}, {Larson}, {Nolta}, {Page}, {Spergel},
  {Halpern}, {Hill}, {Kogut}, {Limon}, {Meyer}, {Odegard}, {Tucker}, {Weiland},
  {Wollack}, \& {Wright}}]{komatsu11}
{Komatsu}, E., {et~al.} 2011, \apjs, 192, 18

\bibitem[{{Lagache} {et~al.}(2005){Lagache}, {Puget}, \& {Dole}}]{lagache05}
{Lagache}, G., {Puget}, J.-L., \& {Dole}, H. 2005, \araa, 43, 727

\bibitem[{{Larson} {et~al.}(2011){Larson}, {Dunkley}, {Hinshaw}, {Komatsu},
  {Nolta}, {Bennett}, {Gold}, {Halpern}, {Hill}, {Jarosik}, {Kogut}, {Limon},
  {Meyer}, {Odegard}, {Page}, {Smith}, {Spergel}, {Tucker}, {Weiland},
  {Wollack}, \& {Wright}}]{larson10}
{Larson}, D., {et~al.} 2011, \apjs, 192, 16

\bibitem[{Lewis {et~al.}(2000)Lewis, Challinor, \& Lasenby}]{lewis99}
Lewis, A., Challinor, A., \& Lasenby, A. 2000, Astrophys. J., 538, 473

\bibitem[{{Lidz} {et~al.}(2007){Lidz}, {McQuinn}, {Zaldarriaga}, {Hernquist},
  \& {Dutta}}]{Lidz2007}
{Lidz}, A., {McQuinn}, M., {Zaldarriaga}, M., {Hernquist}, L., \& {Dutta}, S.
  2007, \apj, 670, 39

\bibitem[{{Lidz} {et~al.}(2006){Lidz}, {Oh}, \& {Furlanetto}}]{LidzOF2006}
{Lidz}, A., {Oh}, S.~P., \& {Furlanetto}, S.~R. 2006, \apjl, 639, L47

\bibitem[{{Malhotra} \& {Rhoads}(2004)}]{MalhotraRhoads2004}
{Malhotra}, S., \& {Rhoads}, J.~E. 2004, \apjl, 617, L5

\bibitem[{{Marsden} {et~al.}(2009){Marsden}, {Ade}, {Bock}, {Chapin}, {Devlin},
  {Dicker}, {Griffin}, {Gundersen}, {Halpern}, {Hargrave}, {Hughes}, {Klein},
  {Mauskopf}, {Magnelli}, {Moncelsi}, {Netterfield}, {Ngo}, {Olmi}, {Pascale},
  {Patanchon}, {Rex}, {Scott}, {Semisch}, {Thomas}, {Truch}, {Tucker},
  {Tucker}, {Viero}, \& {Wiebe}}]{marsden09}
{Marsden}, G., {et~al.} 2009, \apj, 707, 1729

\bibitem[{{Maselli} {et~al.}(2007){Maselli}, {Gallerani}, {Ferrara}, \&
  {Choudhury}}]{Maselli2007}
{Maselli}, A., {Gallerani}, S., {Ferrara}, A., \& {Choudhury}, T.~R. 2007,
  \mnras, 376, L34

\bibitem[{McGreer {et~al.}(2011)McGreer, Mesinger, \& Fan}]{McGreer2011}
McGreer, I.~D., Mesinger, A., \& Fan, X. 2011

\bibitem[{McQuinn {et~al.}(2005{\natexlab{a}})McQuinn, Furlanetto, Hernquist,
  Zahn, \& Zaldarriaga}]{McQuinn2005}
McQuinn, M., Furlanetto, S.~R., Hernquist, L., Zahn, O., \& Zaldarriaga, M.
  2005{\natexlab{a}}, Astrophys. J., 630, 643

\bibitem[{McQuinn {et~al.}(2005{\natexlab{b}})McQuinn, Furlanetto, Hernquist,
  Zahn, \& Zaldarriaga}]{McQuinn:2005ce}
---. 2005{\natexlab{b}}, Astrophys. J., 630, 643

\bibitem[{{McQuinn} {et~al.}(2007{\natexlab{a}}){McQuinn}, {Hernquist},
  {Zaldarriaga}, \& {Dutta}}]{McQuinn2007a}
{McQuinn}, M., {Hernquist}, L., {Zaldarriaga}, M., \& {Dutta}, S.
  2007{\natexlab{a}}, \mnras, 381, 75

\bibitem[{{McQuinn} {et~al.}(2007{\natexlab{b}}){McQuinn}, {Lidz}, {Zahn},
  {Dutta}, {Hernquist}, \& {Zaldarriaga}}]{McQuinn2007b}
{McQuinn}, M., {Lidz}, A., {Zahn}, O., {Dutta}, S., {Hernquist}, L., \&
  {Zaldarriaga}, M. 2007{\natexlab{b}}, \mnras, 377, 1043

\bibitem[{{McQuinn} {et~al.}(2008){McQuinn}, {Lidz}, {Zaldarriaga},
  {Hernquist}, \& {Dutta}}]{McQuinn2008}
{McQuinn}, M., {Lidz}, A., {Zaldarriaga}, M., {Hernquist}, L., \& {Dutta}, S.
  2008, \mnras, 388, 1101

\bibitem[{{Mellema} {et~al.}(2006){Mellema}, {Iliev}, {Alvarez}, \&
  {Shapiro}}]{Mellema2006}
{Mellema}, G., {Iliev}, I.~T., {Alvarez}, M.~A., \& {Shapiro}, P.~R. 2006, New
  Astronomy, 11, 374

\bibitem[{{Mesinger} \& {Furlanetto}(2007)}]{MesingerFurlanetto2007}
{Mesinger}, A., \& {Furlanetto}, S. 2007, \apj, 669, 663

\bibitem[{{Mesinger} \&
  {Furlanetto}(2008{\natexlab{a}})}]{MesingerFurlanetto2008a}
{Mesinger}, A., \& {Furlanetto}, S.~R. 2008{\natexlab{a}}, \mnras, 385, 1348

\bibitem[{{Mesinger} \&
  {Furlanetto}(2008{\natexlab{b}})}]{MesingerFurlanetto2008b}
---. 2008{\natexlab{b}}, \mnras, 386, 1990

\bibitem[{{Mesinger} \& {Haiman}(2004)}]{MesingerHaiman2004}
{Mesinger}, A., \& {Haiman}, Z. 2004, \apjl, 611, L69

\bibitem[{{Mesinger} \& {Haiman}(2007)}]{MesingerHaiman2007}
---. 2007, \apj, 660, 923

\bibitem[{{Mesinger} {et~al.}(2004){Mesinger}, {Haiman}, \&
  {Cen}}]{MesingerHC2004}
{Mesinger}, A., {Haiman}, Z., \& {Cen}, R. 2004, \apj, 613, 23

\bibitem[{Miralda-Escude(2003)}]{MiraldaEscude:2002yd}
Miralda-Escude, J. 2003, Astrophys.J., 597, 66

\bibitem[{Mortlock {et~al.}(2011)}]{Mortlock:2011va}
Mortlock, D.~J., {et~al.} 2011, Nature, 474, 616

\bibitem[{Oh \& Furlanetto(2005)}]{Oh:2004rm}
Oh, S.~P., \& Furlanetto, S.~R. 2005, Astrophys. J., 620, L9

\bibitem[{{Ostriker} \& {Vishniac}(1986)}]{ostriker86}
{Ostriker}, J.~P., \& {Vishniac}, E.~T. 1986, \apjl, 306, L51

\bibitem[{{Ouchi} {et~al.}(2010){Ouchi}, {Shimasaku}, {Furusawa}, {Saito},
  {Yoshida}, {Akiyama}, {Ono}, {Yamada}, {Ota}, {Kashikawa}, {Iye}, {Kodama},
  {Okamura}, {Simpson}, \& {Yoshida}}]{Ouchi2010}
{Ouchi}, M., {et~al.} 2010, \apj, 723, 869

\bibitem[{{Page} {et~al.}(2007){Page}, {Hinshaw}, {Komatsu}, {Nolta},
  {Spergel}, {Bennett}, {Barnes}, {Bean}, {Dor{\'e}}, {Dunkley}, {Halpern},
  {Hill}, {Jarosik}, {Kogut}, {Limon}, {Meyer}, {Odegard}, {Peiris}, {Tucker},
  {Verde}, {Weiland}, {Wollack}, \& {Wright}}]{Page2007}
{Page}, L., {et~al.} 2007, \apjs, 170, 335

\bibitem[{{Parsons} {et~al.}(2010){Parsons}, {Backer}, {Foster}, {Wright},
  {Bradley}, {Gugliucci}, {Parashare}, {Benoit}, {Aguirre}, {Jacobs},
  {Carilli}, {Herne}, {Lynch}, {Manley}, \& {Werthimer}}]{Parsons2009}
{Parsons}, A.~R., {et~al.} 2010, \aj, 139, 1468

\bibitem[{{Pen} {et~al.}(2009){Pen}, {Chang}, {Hirata}, {Peterson}, {Roy},
  {Gupta}, {Odegova}, \& {Sigurdson}}]{Pen2009}
{Pen}, U., {Chang}, T., {Hirata}, C.~M., {Peterson}, J.~B., {Roy}, J., {Gupta},
  Y., {Odegova}, J., \& {Sigurdson}, K. 2009, \mnras, 399, 181

\bibitem[{{Percival} {et~al.}(2010){Percival}, {Reid}, {Eisenstein}, {Bahcall},
  {Budavari}, {Frieman}, {Fukugita}, {Gunn}, {Ivezi{\'c}}, {Knapp}, {Kron},
  {Loveday}, {Lupton}, {McKay}, {Meiksin}, {Nichol}, {Pope}, {Schlegel},
  {Schneider}, {Spergel}, {Stoughton}, {Strauss}, {Szalay}, {Tegmark},
  {Vogeley}, {Weinberg}, {York}, \& {Zehavi}}]{percival10}
{Percival}, W.~J., {et~al.} 2010, \mnras, 401, 2148

\bibitem[{{Petkova} \& {Springel}(2009)}]{PetkovaSpringel2009}
{Petkova}, M., \& {Springel}, V. 2009, \mnras, 396, 1383

\bibitem[{{Planck Collaboration}(2011)}]{planck11-6.6_arxiv}
{Planck Collaboration}. 2011, ArXiv e-prints

\bibitem[{Pritchard \& Loeb(2010)}]{Pritchard:2010pa}
Pritchard, J.~R., \& Loeb, A. 2010, Phys. Rev., D82, 023006

\bibitem[{{Radford} {et~al.}(2007){Radford}, {Giovanelli}, {Sebring}, \&
  {Zmuidzinas}}]{ccat}
{Radford}, S.~J.~E., {Giovanelli}, R., {Sebring}, T.~A., \& {Zmuidzinas}, J.
  2007, in Eighteenth International Symposium on Space Terahertz Technology,
  ed. {A.~Karpov}, 32

\bibitem[{{Razoumov} {et~al.}(2002){Razoumov}, {Norman}, {Abel}, \&
  {Scott}}]{Razoumov2002}
{Razoumov}, A.~O., {Norman}, M.~L., {Abel}, T., \& {Scott}, D. 2002, \apj, 572,
  695

\bibitem[{{Reichardt} {et~al.}(2012){Reichardt}, {Shaw}, {Zahn}, {Aird},
  {Benson}, {Bleem}, {Carlstrom}, {Chang}, {Cho}, {Crawford}, {Crites}, {de
  Haan}, {Dobbs}, {Dudley}, {George}, {Halverson}, {Holder}, {Holzapfel},
  {Hoover}, {Hou}, {Hrubes}, {Joy}, {Keisler}, {Knox}, {Lee}, {Leitch},
  {Lueker}, {Luong-Van}, {McMahon}, {Mehl}, {Meyer}, {Millea}, {Mohr},
  {Montroy}, {Natoli}, {Padin}, {Plagge}, {Pryke}, {Ruhl}, {Schaffer},
  {Shirokoff}, {Spieler}, {Staniszewski}, {Stark}, {Story}, {van Engelen},
  {Vanderlinde}, {Vieira}, \& {Williamson}}]{reichardt11}
{Reichardt}, C.~L., {et~al.} 2012, \apj, 755, 70

\bibitem[{{Riess} {et~al.}(2011){Riess}, {Macri}, {Casertano}, {Lampeitl},
  {Ferguson}, {Filippenko}, {Jha}, {Li}, \& {Chornock}}]{riess11}
{Riess}, A.~G., {et~al.} 2011, \apj, 730, 119

\bibitem[{{Santos} {et~al.}(2003){Santos}, {Cooray}, {Haiman}, {Knox}, \&
  {Ma}}]{Santos:2003jb}
{Santos}, M.~G., {Cooray}, A., {Haiman}, Z., {Knox}, L., \& {Ma}, C. 2003,
  \apj, 598, 756

\bibitem[{{Sehgal} {et~al.}(2010){Sehgal}, {Bode}, {Das},
  {Hernandez-Monteagudo}, {Huffenberger}, {Lin}, {Ostriker}, \&
  {Trac}}]{sehgal10}
{Sehgal}, N., {Bode}, P., {Das}, S., {Hernandez-Monteagudo}, C.,
  {Huffenberger}, K., {Lin}, Y., {Ostriker}, J.~P., \& {Trac}, H. 2010, \apj,
  709, 920

\bibitem[{{Semelin} {et~al.}(2007){Semelin}, {Combes}, \&
  {Baek}}]{SemelinCB2007}
{Semelin}, B., {Combes}, F., \& {Baek}, S. 2007, \aap, 474, 365

\bibitem[{{Shang} {et~al.}(2012){Shang}, {Haiman}, {Knox}, \& {Oh}}]{shang11}
{Shang}, C., {Haiman}, Z., {Knox}, L., \& {Oh}, S.~P. 2012, \mnras, 421, 2832

\bibitem[{{Shaw} {et~al.}(2010){Shaw}, {Nagai}, {Bhattacharya}, \&
  {Lau}}]{shaw10}
{Shaw}, L.~D., {Nagai}, D., {Bhattacharya}, S., \& {Lau}, E.~T. 2010, \apj,
  725, 1452

\bibitem[{{Shaw} {et~al.}(2011){Shaw}, {Rudd}, \& {Nagai}}]{shaw11}
{Shaw}, L.~D., {Rudd}, D.~H., \& {Nagai}, D. 2011, \apj, submitted,
  arXiv:1109.0553

\bibitem[{{Shaw} {et~al.}(2009){Shaw}, {Zahn}, {Holder}, \&
  {Dor{\'e}}}]{shaw09}
{Shaw}, L.~D., {Zahn}, O., {Holder}, G.~P., \& {Dor{\'e}}, O. 2009, \apj, 702,
  368

\bibitem[{{Shin} {et~al.}(2008){Shin}, {Trac}, \& {Cen}}]{ShinTC2008}
{Shin}, M.-S., {Trac}, H., \& {Cen}, R. 2008, \apj, 681, 756

\bibitem[{{Sokasian} {et~al.}(2001){Sokasian}, {Abel}, \&
  {Hernquist}}]{Sokasian2001}
{Sokasian}, A., {Abel}, T., \& {Hernquist}, L.~E. 2001, New Astronomy, 6, 359

\bibitem[{Sunyaev \& Zel'dovich(1980)}]{sunyaev80}
Sunyaev, R., \& Zel'dovich, Y. 1980, ARAA, 18, 537

\bibitem[{{Sunyaev} \& {Zel'dovich}(1970)}]{sunyaev70b}
{Sunyaev}, R.~A., \& {Zel'dovich}, Y.~B. 1970, \apss, 7, 3

\bibitem[{{Thomas} {et~al.}(2009){Thomas}, {Zaroubi}, {Ciardi}, {Pawlik},
  {Labropoulos}, {Jeli{\'c}}, {Bernardi}, {Brentjens}, {de Bruyn}, {Harker},
  {Koopmans}, {Mellema}, {Pandey}, {Schaye}, \& {Yatawatta}}]{Thomas2009}
{Thomas}, R.~M., {et~al.} 2009, \mnras, 393, 32

\bibitem[{{Tinker} {et~al.}(2008){Tinker}, {Kravtsov}, {Klypin}, {Abazajian},
  {Warren}, {Yepes}, {Gottl{\"o}ber}, \& {Holz}}]{tinker08}
{Tinker}, J., {Kravtsov}, A.~V., {Klypin}, A., {Abazajian}, K., {Warren}, M.,
  {Yepes}, G., {Gottl{\"o}ber}, S., \& {Holz}, D.~E. 2008, \apj, 688, 709

\bibitem[{{Totani} {et~al.}(2006){Totani}, {Kawai}, {Kosugi}, {Aoki}, {Yamada},
  {Iye}, {Ohta}, \& {Hattori}}]{Totani2006}
{Totani}, T., {Kawai}, N., {Kosugi}, G., {Aoki}, K., {Yamada}, T., {Iye}, M.,
  {Ohta}, K., \& {Hattori}, T. 2006, \pasj, 58, 485

\bibitem[{{Trac} {et~al.}(2011){Trac}, {Bode}, \& {Ostriker}}]{trac11}
{Trac}, H., {Bode}, P., \& {Ostriker}, J.~P. 2011, \apj, 727, 94

\bibitem[{{Trac} \& {Cen}(2007{\natexlab{a}})}]{TracCen2007}
{Trac}, H., \& {Cen}, R. 2007{\natexlab{a}}, \apj, 671, 1

\bibitem[{{Trac} \& {Cen}(2007{\natexlab{b}})}]{trac07}
---. 2007{\natexlab{b}}, \apj, 671, 1

\bibitem[{{Trac} {et~al.}(2008){Trac}, {Cen}, \& {Loeb}}]{TracCL2008}
{Trac}, H., {Cen}, R., \& {Loeb}, A. 2008, \apjl, 689, L81

\bibitem[{{Trac} \& {Gnedin}(2011)}]{TracGnedin2009}
{Trac}, H.~Y., \& {Gnedin}, N.~Y. 2011, Advanced Science Letters, 4, 228

\bibitem[{{Wetzel} \& {White}(2010)}]{wetzel10}
{Wetzel}, A.~R., \& {White}, M. 2010, \mnras, 403, 1072

\bibitem[{{Wootten} \& {Thompson}(2009)}]{alma}
{Wootten}, A., \& {Thompson}, A.~R. 2009, IEEE Proceedings, 97, 1463

\bibitem[{{Wyithe} \& {Cen}(2007)}]{WyitheCen2007}
{Wyithe}, J.~S.~B., \& {Cen}, R. 2007, \apj, 659, 890

\bibitem[{{Wyithe} {et~al.}(2005){Wyithe}, {Loeb}, \& {Carilli}}]{WyitheLC2005}
{Wyithe}, J.~S.~B., {Loeb}, A., \& {Carilli}, C. 2005, \apj, 628, 575

\bibitem[{{Zahn} {et~al.}(2007){Zahn}, {Lidz}, {McQuinn}, {Dutta}, {Hernquist},
  {Zaldarriaga}, \& {Furlanetto}}]{Zahn2007}
{Zahn}, O., {Lidz}, A., {McQuinn}, M., {Dutta}, S., {Hernquist}, L.,
  {Zaldarriaga}, M., \& {Furlanetto}, S.~R. 2007, \apj, 654, 12

\bibitem[{{Zahn} {et~al.}(2011){Zahn}, {Mesinger}, {McQuinn}, {Trac}, {Cen}, \&
  {Hernquist}}]{Zahn2010}
{Zahn}, O., {Mesinger}, A., {McQuinn}, M., {Trac}, H., {Cen}, R., \&
  {Hernquist}, L.~E. 2011, \mnras, 414, 727

\bibitem[{{Zahn} {et~al.}(2005{\natexlab{a}}){Zahn}, {Zaldarriaga},
  {Hernquist}, \& {McQuinn}}]{zahn05}
{Zahn}, O., {Zaldarriaga}, M., {Hernquist}, L., \& {McQuinn}, M.
  2005{\natexlab{a}}, \apj, 630, 657

\bibitem[{{Zahn} {et~al.}(2005{\natexlab{b}}){Zahn}, {Zaldarriaga},
  {Hernquist}, \& {McQuinn}}]{Zahn2005}
---. 2005{\natexlab{b}}, \apj, 630, 657

\bibitem[{{Zaldarriaga} {et~al.}(2004){Zaldarriaga}, {Furlanetto}, \&
  {Hernquist}}]{ZaldarriagaFH2004}
{Zaldarriaga}, M., {Furlanetto}, S.~R., \& {Hernquist}, L. 2004, \apj, 608, 622

\bibitem[{Zaldarriaga {et~al.}(2008)}]{Zaldarriaga:2008ap}
Zaldarriaga, M., {et~al.} 2008

\end{thebibliography}
